\documentstyle[prd,aps,epsf]{revtex}

\def\simgt{\,\rlap{\lower 3.5 pt\hbox{$\mathchar \sim$}}\raise 1pt \hbox {$>$}\,}
\def\simlt{\,\rlap{\lower 3.5 pt\hbox{$\mathchar \sim$}}\raise 1pt \hbox {$<$}\,}
\begin{document}

\draft

\tightenlines

\title{
\vspace*{-35pt}
{\normalsize \hfill {\sf SWAT/02/335}} \\ 
{\normalsize \hfill {\sf NSF-ITP-02-26}} \\ 
{\normalsize \hfill {\sf BI-TP 2002/06}} \\ 
The QCD thermal phase transition in the presence of a 
small chemical potential
}

\author{
C.R. Allton\rlap,$^{\rm a,b}$ S. Ejiri\rlap,$^{\rm a}$
S.J. Hands\rlap,$^{\rm a,c}$ O. Kaczmarek\rlap,$^{\rm d}$
F. Karsch\rlap,$^{\rm c,d}$ E. Laermann\rlap,$^{\rm d}$
Ch. Schmidt\rlap,$^{\rm d}$ and L. Scorzato$^{\rm a}$
\footnote{Present address: DESY Theory Division, Notkestrasse 85,
D-22603 Hamburg, Germany.}}

\address{
$^{\rm a}$Department of Physics, University of Wales Swansea,
          Singleton Park, Swansea, SA2 8PP, U.K. \\
$^{\rm b}$Department of Mathematics, University of Queensland,
Brisbane 4072, Australia\\ 
$^{\rm c}$ Institute for Theoretical Physics, University of 
California, Santa Barbara, CA 93106-4030, U.S.A.\\
$^{\rm d}$Fakult\"at f\"ur Physik, Universit\"at Bielefeld,
          D-33615 Bielefeld, Germany} 

\date{\today}
\maketitle

\begin{abstract}
We propose a new method to investigate the thermal properties of QCD 
with a small quark chemical potential $\mu$. 
Derivatives of quark and gluonic observables with respect to $\mu$ 
are computed at $\mu=0$ for 2 flavors of p-4 improved staggered fermions
with $ma=0.1,0.2$
on a $16^3\times4$ lattice, 
and used to calculate the leading order Taylor expansion in $\mu$ of the 
location of the pseudocritical point about $\mu=0$.
This expansion should be well-behaved for 
the small values of $\mu_{\rm q}/T_c\sim0.1$ relevant for RHIC phenomenology, 
and 
predicts a critical curve $T_c(\mu)$ in reasonable agreement with estimates
obtained using exact reweighting. In addition, 
we contrast the case of isoscalar and isovector chemical
potentials,  quantify the effect of $\mu\not=0$ 
on the equation of state, and 
comment on the complex phase of the fermion determinant in QCD with $\mu\not=0$.
\end{abstract}




\section{Introduction}
\label{sec:intro}

The study of the phase structure of QCD at non-zero temperature and baryon
density 
is one of the most interesting topics in contemporary physics.
Heavy-ion collision experiments are running at BNL and CERN with the goal of
the experimental production of a new state of matter, the quark-gluon plasma
\cite{Heinz00}. 
On the theoretical side, novel
color superconducting and superfluid phases have been 
conjectured at high baryon densities \cite{CSC}. 
For these reasons the need for numerical studies of the QCD phase transition 
using lattice gauge theory simulations,
currently the most powerful quantitative approach to QCD,
with both 
temperature $T\not=0$ and quark chemical potential $\mu_{\rm q}\not=0$, is
more urgent than ever.
Precise theoretical inputs from simulations in the vicinity of 
the QCD phase transition 
are indispensable to the understanding of the heavy-ion collision experiments. 

Over the last several years, the numerical study of lattice QCD 
has been successful at zero chemical potential and 
high temperature\cite{Eji00}.
In contrast, because the quark determinant is complex at 
$\mu \neq 0$ and Monte-Carlo simulation not directly applicable, 
studies at non-zero $\mu$ are still largely exploratory.
Recent developments with $\mu\not=0$ can be classified in two categories
\cite{Hands01}.
At the low temperatures and high densities where the new phases are 
expected,
studies of model field theories 
such as two-color (SU(2)) QCD and the NJL model have been made. 
The simulation is possible 
because in both cases the quark determinant is 
positive definite so that conventional
Monte Carlo methods can be used.
The other case is high temperature and low density, 
which is phenomenologically more important for RHIC, 
since the QCD phase transition 
both in the early universe 
and in the interesting regime for heavy-ion collisions 
is expected at rather low density, 
e.g., $\mu_{\rm q} \sim 15 {\rm MeV}$ $(\mu_{\rm q}/T_c \sim0.1)$ 
for RHIC \cite{Redl01}. 
In this region the reweighting method, in which observables
at $\mu\not=0$  are computed by performing simulations 
at ${\rm Re} (\mu) =0$, is applicable \cite{Glasgow}.
Using this method, the first results on the phase structure 
in the $(\mu,T)$ plane were recently obtained by Fodor and Katz
\cite{Fod01}. 
Unfortunately, although in principle with infinite statistics this method is
exact, rather general arguments suggest that in practise
the region of applicability of the reweighting method becomes 
narrower as the lattice volume is increased.
Another efficient method at low density is via a Taylor 
expansion obtained by computing the derivatives 
of physical quantities with respect to $\mu$ at $\mu=0$.  
This approach is not restricted to small lattices, because
it requires only the expectation values of local fermion bilinears;
these are measured effectively on large systems using stochastic methods, 
and might 
even be expected to
self-average as the volume increases.
Since analyticity is required, however, 
the values of $\mu$ which can be reached
must be bounded 
by, e.g.  
the critical point expected in the $(\mu,T)$ plane for QCD with 2
light flavors.
Pioneering work in such a framework have been made by developing expansions
for free energy, yielding 
quark number susceptibility~\cite{Gott88,Gav01}, for  
hadronic screening masses\cite{Taro01} and 
in the context of the 3-dimensional effective theory\cite{Hart01}.

In this study, we investigate the transition temperature $T_c$ as
a function of $\mu\not=0$.
In Sec.~\ref{sec:method}, we propose a new method to compute 
derivatives of physical quantities with respect to $\mu.$ 
Details of our simulations performed on a $16^3\times4$ lattice with 
quark masses $m=0.1,0.2$ are presented in Sec.~\ref{sec:simulation}.
In Sec.~\ref{sec:mass} we check the feasibility of the method by
calculating the derivative of the transition point with respect to 
$m$.
Our main result, the calculation of
the second derivative of $\beta_c$ with respect to $\mu$ for 2 flavor QCD, 
is given in Sec.~\ref{sec:mu}. Using data on the lattice beta-function, we are 
then able to translate this result into 
physical units, yielding an estimate for the phase transition line $T_c(\mu)$.
We also discuss the response of the pressure $p(T)$ and energy density 
$\epsilon(T)$ to non-zero $\mu$, and estimate their variation along the 
critical line. Finally in this section we discuss the 
problem of the complex phase of the quark 
determinant, and show that the sign problem is mild in the region of the 
phase diagram relevant for RHIC physics.
Section \ref{sec:summary} presents our conclusions.

\section{Reweighting method for the $\mu$-direction}
\label{sec:method}

Ferrenberg and Swendsen's reweighting method is a very useful technique 
to investigate critical phenomena \cite{Swen88}. 
In QCD the expectation value of an observable ${\cal O}(\beta, m, \mu)$ 
can in principle be computed by
simulation at $(\beta_0, m_0, \mu_0)$ using the following identity:
\begin{eqnarray}
\langle {\cal O} \rangle_{(\beta, m, \mu)}
&=& \frac{1}{{\cal Z}(\beta, m, \mu)} \int 
{\cal D}U {\cal O} ( \det M(m, \mu))^{\alpha N_{\rm f}} {\rm e}^{-S_g(\beta)} 
\label{eq:rew} \\
&=& \frac{\left\langle {\cal O} {\rm e}^{\alpha N_{\rm f}
(\ln \det M(m, \mu)-\ln \det M(m_0,\mu_0))} {\rm e}^{-S_g(\beta)+S_g(\beta_0)}
 \right\rangle_{(\beta_0, m_0, \mu_0)} }{
\left\langle{\rm e}^{\alpha N_{\rm f}
(\ln \det M(m, \mu)-\ln \det M(m_0,\mu_0))} {\rm e}^{-S_g(\beta)+S_g(\beta_0)}
  \right\rangle_{(\beta_0, m_0, \mu_0)}}\label{eq:opexp}.
\end{eqnarray}
Here $M$ is the quark matrix, $S_g$ the gauge action, 
$N_{\rm f}$ the number of flavors, and $\alpha=1$ or $1/4$ 
for Wilson or staggered lattice fermions respectively.
The chemical potential parameter $\mu=\mu_{\rm q}a$, where $a$ is the
lattice spacing.
Because $\det M(\mu)$ is complex for Re$(\mu)\not=0$, the expectation values in 
(\ref{eq:opexp}) can only be estimated by conventional Monte Carlo importance
sampling if the simulation is performed for $\mu_0$ zero or pure imaginary.
Most of the attempts to calculate at $\mu\not=0$ have used 
variants of this method \cite{Glasgow}.
The reweighting factor for the gauge part is easy to compute by 
measuring the plaquette $P_{\mu\nu}$, since
\begin{eqnarray}
-S_g(\beta)+S_g(\beta_0)=
(\beta-\beta_0) \sum_{x, \mu > \nu} P_{\mu \nu} (x),
\end{eqnarray}
for the standard 
Wilson action, and extensions for improved actions are easy to derive. 
However, to compute the fermion part, the calculation of the fermion 
determinant is required for each point $(m, \mu)$ we want to study. 
Such a calculation is quite expensive and difficult to perform in 
practice. Fodor and Katz have performed such calculations, 
and by reweighting in both $\mu$ and $\beta$ have succeeded
in tracing out the critical line $\beta_c(\mu)$ and locating the 
critical endpoint on small lattices \cite{Fod01}.
Their method exploits the fact that the overlap between
ensembles at different points 
along the coexistence line
separating hadronic and quark-gluon plasma phases remains reasonably large
on finite
systems.

Another problem of the reweighting method is the sign problem, 
which will be discussed in detail later. 
As $\mu$ increases from zero, the calculation of eqn.(\ref{eq:opexp}) 
becomes more difficult due to fluctuations in the phase of the
denominator. 
To avoid these problems, we restrict ourselves to calculating derivatives 
of physical quantities with respect to $\mu$, 
which can be done at $\mu=0$. 
This yields estimates of the physical quantity 
as a continuous function of $\mu$ in a narrow range of $\mu$, but 
the region of applicability is 
not restricted to the immediate neighbourhood of the phase transition. 
This permits the development of a Taylor expansion of observables in powers
of $\mu=\mu_{\rm q}a$; strictly speaking, in fact, the physically relevant
expansion parameter which ultimately must govern convergence is the 
fugacity $\mu_{\rm q}/T=N_t\mu$.
The Taylor expansion
for the 
fermionic part of the reweighting factor around $\mu=0$ is
\begin{eqnarray}
\label{eq:expand}
\alpha N_{\rm f}\ln\left({{\det M(\mu)}\over{\det M(0)}}\right) = 
\alpha N_{\rm f}\sum_{n=1}^{\infty} \frac{\mu^n}{n!} 
\frac{\partial^n \ln \det M(0)}{\partial \mu^n}\equiv\sum_{n=1}^\infty
{\cal R}_n\mu^n.
\end{eqnarray}
We similarly expand fermionic observables such as 
the chiral condensate,
\begin{eqnarray}
\langle \bar{\psi} \psi \rangle 
= (N_s^3 \times N_t)^{-1} \alpha N_{\rm f} \langle {\rm tr} M^{-1} \rangle,
\end{eqnarray}
where the lattice size is $N_s^3 \times N_t$,
once again obtaining a continuous function for small $\mu$.
Using the formula 
\begin{equation}
{{\partial M}\over{\partial x}}^{-1}=-M^{-1}{{\partial M}\over{\partial x}}
M^{-1},
\end{equation}
expressions for $\partial^n (\ln \det M) / \partial \mu^n$ and 
$\partial^n ({\rm tr} M^{-1}) / \partial \mu^n$ in terms of traces over
products
of local operators and inverse matrices can be developed:
\begin{eqnarray}
\frac{\partial \ln \det M}{\partial \mu} 
&=& {\rm tr} \left( M^{-1} \frac{\partial M}{\partial \mu} \right), \nonumber\\
\frac{\partial^2 \ln \det M}{\partial \mu^2} 
&=& {\rm tr} \left( M^{-1} \frac{\partial^2 M}{\partial \mu^2} \right)
 - {\rm tr} \left( M^{-1} \frac{\partial M}{\partial \mu}
                   M^{-1} \frac{\partial M}{\partial \mu} \right), \nonumber\\
\label{eq:dermu}
\frac{\partial^3 \ln \det M}{\partial \mu^3} 
&=& {\rm tr} \left( M^{-1} \frac{\partial^3 M}{\partial \mu^3} \right)
 -3 {\rm tr} \left( M^{-1} \frac{\partial M}{\partial \mu}
              M^{-1} \frac{\partial^2 M}{\partial \mu^2} \right)
 +2 {\rm tr} \left( M^{-1} \frac{\partial M}{\partial \mu}
        M^{-1} \frac{\partial M}{\partial \mu}
        M^{-1} \frac{\partial M}{\partial \mu} \right), 
\end{eqnarray}
\begin{eqnarray}
{{\partial {\rm tr} M^{-1}}\over{\partial \mu}} 
&=&  - {\rm tr} \left( M^{-1} \frac{\partial M}{\partial \mu}
 M^{-1} \right), \nonumber\\
{{\partial^2 {\rm tr} M^{-1}}\over{\partial \mu^2}}
&=& - {\rm tr} \left( M^{-1} \frac{\partial^2 M}{\partial \mu^2}
 M^{-1} \right)
 + 2 {\rm tr} \left( M^{-1} \frac{\partial M}{\partial \mu}
    M^{-1} \frac{\partial M}{\partial \mu} M^{-1} \right), \nonumber\\
{{\partial^3 {\rm tr} M^{-1}}\over{\partial \mu^3}}
&=& - {\rm tr} \left( M^{-1} \frac{\partial^3 M}{\partial \mu^3}
 M^{-1} \right)
 +3 {\rm tr} \left( M^{-1} \frac{\partial^2 M}{\partial \mu^2}
    M^{-1} \frac{\partial M}{\partial \mu} M^{-1} \right), \nonumber \\
&& +3 {\rm tr} \left( M^{-1} \frac{\partial M}{\partial \mu}
    M^{-1} \frac{\partial^2 M}{\partial \mu^2} M^{-1} \right)
-6 {\rm tr} \left( M^{-1} \frac{\partial M}{\partial \mu}
    M^{-1} \frac{\partial M}{\partial \mu} M^{-1} 
    \frac{\partial M}{\partial \mu} M^{-1} \right).
\label{eq:taypbp}
\end{eqnarray}

We apply the random noise method to calculate the derivatives of 
$\ln \det M$ and ${\rm tr} M^{-1}$, 
which enables us to compute on rather large volumes 
in comparison with usual studies of QCD with $\mu\not=0$.
Using $N_{\rm n}$ sets of random noise vectors $\eta_{ai}$ 
which satisfy the condition: 
$\lim_{N_{\rm n} \rightarrow \infty} (1/N_{\rm n}) 
\sum_{a=1}^{N_{\rm n}} \eta^{*}_{ai} \eta_{aj} = \delta_{ij},$
we rewrite the trace of products of 
$\partial M / \partial \mu$ and $M^{-1}$ as
\begin{eqnarray}
\label{eq:noise}
{\rm tr} \left( \frac{\partial^{n_1} M}{\partial \mu^{n_1}} M^{-1} 
\frac{\partial^{n_2} M}{\partial \mu^{n_2}} \cdots M^{-1} \right) = 
\lim_{N_{\rm n} \rightarrow \infty} \frac{1}{N_{\rm n}} 
\sum_{a=1}^{N_{\rm n}} \eta^{\dagger}_{a} 
\frac{\partial^{n_1} M}{\partial \mu^{n_1}} M^{-1} 
\frac{\partial^{n_2} M}{\partial \mu^{n_2}} \cdots M^{-1} \eta_{a}.
\end{eqnarray}
$M^{-1} \eta_{a} \equiv x$ and 
$M^{-1} (\partial M/ \partial \mu) \cdots \eta_{a} \equiv x$ 
are obtained by solving $M x = \eta_{a}$ or 
$M x = (\partial M/ \partial \mu) \cdots \eta_{a}$,  
and we compute the RHS of eqn.(\ref{eq:noise}) with finite $N_{\rm n}$.
The error for estimates of physical observables made from 
$N_{\rm conf}$ configurations 
is expected to decrease as
$(N_{\rm n}N_{\rm conf})^{-1/2}$. Further notes on 
the application of the noise method are given in the Appendix.

By using the derivatives of both the reweighting 
factor and fermionic observable up to $n$-th order in $\mu$, 
we can obtain the correct answer for the expectation value 
up to $n$-th order, which can be easily checked by performing a Taylor 
expansion of the expectation value, eqn.(\ref{eq:rew}), directly 
for each physical observable. 
Of course, for a pure gluonic observable such as
the Polyakov loop $L$ only the expansion of $\ln\det M$ is needed.
Furthermore, we should note that at $\mu=0$ the odd order derivatives of 
both $\ln \det M$ and ${\rm tr} M^{-1}$ are pure imaginary and 
the even order derivatives are real.
This property is proved using the identities for the fermion matrix:
\begin{eqnarray}
M^{\dagger}(\mu) = \Gamma_5 M(-\mu) \Gamma_5,
\hspace{5mm} {\rm and} \hspace{5mm}
\frac{\partial^n M^{\dagger}}{\partial \mu^n}(\mu) = 
(-1)^n \Gamma_5 \frac{\partial^n M}{\partial \mu^n}(-\mu) \Gamma_5,
\end{eqnarray}
where $\Gamma_5$ is $\gamma_5$ for Wilson fermions and 
$(-1)^{x_1+x_2+x_3+x_4}$ for staggered.
Then, at $\mu=0$
\begin{eqnarray}
{\rm tr} \left( M^{-1} \frac{\partial^{n_1} M}{\partial \mu^{n_1}}
    M^{-1} \frac{\partial^{n_2} M}{\partial \mu^{n_2}} M^{-1} 
    \cdots \right)^{*} = (-1)^{n_1+n_2+\cdots}
{\rm tr} \left( M^{-1} \frac{\partial^{n_1} M}{\partial \mu^{n_1}}
    M^{-1} \frac{\partial^{n_2} M}{\partial \mu^{n_2}} M^{-1} 
    \cdots \right).
\end{eqnarray}
Because the terms in the $n$-th derivative satisfy 
$n_1 + n_2 + \cdots = n$, we obtain
\begin{eqnarray}
\left( \frac{\partial^n \ln \det M}{\partial \mu^n} \right)^* 
&=& (-1)^n \frac{\partial^n \ln \det M}{\partial \mu^n}; \\
\left( \frac{\partial^n {\rm tr} M^{-1}}{\partial \mu^n} \right)^* 
&=& (-1)^n \frac{\partial^n {\rm tr} M^{-1}}{\partial \mu^n}.
\end{eqnarray}
Using this property and the fact that ${\cal Z}$ is a real function of 
$\beta$, $m$ and $\mu$, we can explicitly confirm that, 
if the operator has the property such that even order derivatives are real 
and odd order derivatives are pure imaginary at $\mu=0$, e.g., 
$\langle \bar{\psi} \psi \rangle$ or its susceptibility, 
then all odd order derivatives of the expectation value of a physical 
quantity are zero at $\mu=0$, 
as we expect from the symmetry under changing $\mu$ to $-\mu$. 
The derivative of the expectation value can be written as 
a sum of products of expectation values composed of the operator, 
the reweighting factor and their derivatives, and
the total number of differentiations in each term has to be odd
for an odd order derivative.
Hence all terms for odd derivatives contain at least one 
expectation value of a pure imaginary operator and hence
vanish, since the expectation value of 
a pure imaginary operator is zero.
Therefore the first non-trivial order of corrections to e.g.
$\langle\bar\psi\psi\rangle$ or its susceptbility, which we compute in this 
study is $O(\mu^2)$; the truncation errors, so far unquantified, 
are $O(\mu^4)$.

In order to be more specific, let us define the Taylor
expansion of an operator by $\sum_{n=0}^\infty{\cal O}_n\mu^n$. Then to
$O(\mu^2)$ the
expression (\ref{eq:opexp}) for $\langle{\cal O}\rangle_{(\beta,\mu)}$ can be
rewritten 
\begin{equation}
\langle{\cal O}\rangle_{(\beta,\mu)}=
{
{\langle({\cal O}_0+
 {\cal O}_1\mu+{\cal O}_2\mu^2)
  \exp({\cal R}_1\mu+{\cal R}_2\mu^2-\Delta S_g)\rangle}
\over
{\langle\exp({\cal R}_1\mu+{\cal R}_2\mu^2-\Delta S_g)
\rangle}},
\label{eq:specific}
\end{equation}
where 
expectation values on the RHS 
are measured with respect to an ensemble generated at $(\beta_0,0)$. 
Extension of this formula to combining data from several ensembles using 
multi-histogramming is straightforward \cite{Swen88}. Further details on the 
evaluation of (\ref{eq:specific}) using the noise method for fermionic operators
are given in the Appendix.

In order to determine the pseudocritical point, we calculate the Polyakov 
loop susceptibility, 
\begin{eqnarray} 
\chi_L = N_s^3 \left( \langle L^2 \rangle - \langle L \rangle^2 \right),
\end{eqnarray}
where the Polyakov loop $L= (N_s^3)^{-1} 
\sum_{\vec{x}} N_c^{-1} {\rm tr} \prod_t U_4 (\vec{x},t) $,
and the susceptibility of the chiral condensate\footnote{
Note that we only calculate the disconnected part of 
the complete chiral susceptibility.},
\begin{eqnarray} 
\chi_{\bar{\psi}\psi} = (N_s^3 \times N_t)^{-1} 
(\alpha N_{\rm f})^2 \left( \langle ({\rm tr} M^{-1})^2 \rangle - 
\langle {\rm tr} M^{-1} \rangle^2 \right).
\end{eqnarray}
We define the transition point $\beta_c(\mu)$ by the peak position of these 
susceptibilities for each $\mu$;
\begin{equation}
 {{\partial\chi(\beta_c,\mu)}\over{\partial\beta}} = 0. 
\label{eq:pseudo}
\end{equation}
If we compute $\partial \chi / \partial \beta$ correctly up to $n$-th order 
in $\mu$, we can determine the $n$-th derivative of $\beta_c$ with 
respect to $\mu$. For example 
if we determine $\beta_c(\mu)$ using an operator such as 
$\langle\bar\psi\psi\rangle$, which is real and whose first derivative at
mu=0 is pure imaginary, then the first derivative $\beta_c^\prime(\mu)$ 
vanishes because as argued above the first derivative of 
the susceptibility is zero in this case.

Finally, note we can also estimate the magnitude of fluctuations 
of the phase of $\det M$,
because on each configuration this 
phase can be expressed in terms of the odd terms
of the Taylor expansion of $\ln \det M$; 
this will be discussed in more detail in section \ref{sec:phase}.

\section{Simulations for $N_{\rm f}=2$ improved staggered fermions}
\label{sec:simulation}

We employ a combination of the Symanzik improved gauge and 
2 flavors of the p4-improved staggered fermion actions
\cite{Hel99,Kar00}. 
The partition function is defined by
\begin{eqnarray}
{\cal Z}(\beta, m, \mu) &=& \int 
{\cal D}U (\det M)^{N_{\rm f}/4} {\rm e}^{-S_g},
\label{eq:partition} \\
 S_g & = & -\beta \left\{ \sum_{x,\, \mu > \nu} 
      c_{0} W_{\mu \nu}^{1 \times 1}(x) 
    + \sum_{x,\, \mu, \nu} c_{1} W_{\mu \nu}^{1 \times 2}(x) \right\},
\label{eq:gaction} \\
  M_{x,y}&=&\sum_{i} \eta_{i}(x) \biggl\{
  c_{1}^{\rm F}\Bigl[U^{\rm fat}_i(x)~\delta_{x+\hat{i},y}
    -{U_i^{\rm fat}}^\dagger(x-\hat{i})~\delta_{x-\hat{i},y}\Bigr]
                                         \nonumber \\
  &&  +\;\;\;\;\;\;\;\;\;\;\;\;\;\;c_{3}^{\rm F}\sum_{i \neq j}\Bigl[
    U^{(1,2)}_{i,j}(x)~\delta_{x+\hat{i}+2\hat{j},y}
    - {U^{(1,2)}_{i,j}}^\dagger(x-\hat{i}-2\hat{j})
      \delta_{x-\hat{i}-2\hat{j},y} \nonumber  \\
  && +\;\;\;\;\;\;\;\;\;\;\;\;\;\;\;\;\;\;\;\;\;\;\;\;\;\;
U^{(1,-2)}_{i,j}(x)~\delta_{x+\hat{i}-2\hat{j},y}
    - {U^{(1,-2)}_{i,j}}^\dagger(x-\hat{i}+2\hat{j}) 
     \delta_{x-\hat{i}+2\hat{j},y} \Bigr] \nonumber  \\
  &&  +\;\;\;\;\;\;\;\;\;\;\;\;\;\;c_{3}^{\rm F} \Bigl[
    {\rm e}^{2\mu} U^{(1,2)}_{i,4}(x)~\delta_{x+\hat{i}+2\hat{4},y}
    - {\rm e}^{-2\mu} {U^{(1,2)}_{i,4}}^\dagger(x-\hat{i}-2\hat{4})
      \delta_{x-\hat{i}-2\hat{4},y} \nonumber  \\
  && +\;\;\;\;\;\;\;\;\;\;\;\;\;\;\;\;\;
    {\rm e}^{-2\mu} U^{(1,-2)}_{i,4}(x)~\delta_{x+\hat{i}-2\hat{4},y}
    - {\rm e}^{2\mu} {U^{(1,-2)}_{i,4}}^\dagger(x-\hat{i}+2\hat{4}) 
     \delta_{x-\hat{i}+2\hat{4},y} \Bigr] \biggr\} \nonumber  \\
  && + \eta_{4}(x) \biggl\{
  c_{1}^{\rm F}\Bigr[{\rm e}^{\mu} U^{\rm fat}_4(x)~\delta_{x+\hat{4},y}
    -{{\rm e}^{-\mu} U_4^{\rm fat}}^\dagger(x-\hat{4})~\delta_{x-\hat{4},y}
    \Bigr] \nonumber \\
  &&  +\;\;\;\;\;\;\;\;\;\;\;c_{3}^{\rm F} \sum_{i}\Bigl[
    {\rm e}^{\mu} U^{(1,2)}_{4,i}(x)~\delta_{x+\hat{4}+2\hat{i},y}
    - {\rm e}^{-\mu} {U^{(1,2)}_{4,i}}^\dagger(x-\hat{4}-2\hat{i})
      \delta_{x-\hat{4}-2\hat{i},y}  \nonumber  \\
  && +\;\;\;\;\;\;\;\;\;\;\;\;\;\;\;\;\;\;\;\;\;\;\;
      {\rm e}^{\mu} U^{(1,-2)}_{4,i}(x)~\delta_{x+\hat{4}-2\hat{i},y}
    - {\rm e}^{-\mu} {U^{(1,-2)}_{4,i}}^\dagger(x-\hat{4}+2\hat{i}) 
     \delta_{x-\hat{4}+2\hat{i},y} \Bigr] \biggr\} 
   + m \delta_{x,y}, \label{eq:fermact}
\end{eqnarray}
where $W_{\mu \nu}^{1 \times 1}(x)$ and $W_{\mu \nu}^{1 \times 2}(x)$ are 
$1 \times 1$ and $1 \times 2$ Wilson loops, 
$\eta_\mu(x)=(-1)^{x_1+\cdots+x_{\mu-1}}$ is the KS phase and
\begin{eqnarray}
  U^{(1,2)}_{\mu,\nu}(x) &=& \frac{1}{2} \left[U_\mu(x)
    U_\nu(x+\hat{\mu})U_\nu(x+\hat{\mu}+\hat{\nu})
    + U_\nu(x)U_\nu(x+\hat{\nu})U_\mu(x+2\hat{\nu})\right]~, \nonumber \\
  U^{(1,-2)}_{\mu,\nu}(x) &=& \frac{1}{2} \left[U_\mu(x)
    U^\dagger_\nu(x+\hat{\mu}-\hat{\nu})
    U^\dagger_\nu(x+\hat{\mu}-2\hat{\nu}) + U^\dagger_\nu(x-\hat{\nu})
  U^\dagger_\nu(x-2\hat{\nu})U_\mu(x-2\hat{\nu}) \right], \label{fermact2}\\
  U^{\rm fat}_\mu(x) &=& \frac{1}{1 + 6 \omega} \biggl\{
  U_\mu(x) + \omega \sum_{\nu \neq \mu} \left[
   U_\nu(x)U_\mu(x+\hat{\nu})U^\dagger_\nu(x+\hat{\mu})
    + U^\dagger_\nu(x-\hat{\nu})U_\mu(x-\hat{\nu})U_\nu(x+\hat{\mu}-\hat{\nu})
  \right] \biggr\}. \nonumber
\end{eqnarray}
The coefficients are 
$\beta=6/g^2$, $c_1=-1/12$, $c_0=1-8c_1$, $c_1^{\rm F}=3/8$, 
$c_3^{\rm F}=1/96$, and $\omega=0.2$.
The action is derived such that rotational invariance of the free fermion 
propagator is restored up to $O(p^4)$.
It is known that this action makes the discretization error of 
the equation of state pressure $p(T)$
small as $T\to\infty$, and $T_c$ obtained by this action is consistent 
with that obtained using improved Wilson fermions\cite{Kar00,CPPACS00}.
To incorporate the chemical potential, we generalise the standard
prescription of treating $\mu$
as an imaginary gauge potential $A_0$ \cite{HasKarsch}
by multiplying the terms 
generating $n$-step hops in the positive and negative temporal directions 
by ${\rm e}^{n \mu}$ and ${\rm e}^{-n \mu}$ respectively
\footnote{
Note that for any improved action involving terms in which $\psi$ and $\bar\psi$
are separated by more than a single link, there is no longer a local conserved 
baryon number current bilinear $j_\mu(x)$ such that $\sum_\mu\langle j_\mu(x)
-j_\mu(x-\hat\mu)\rangle=0$ for
non-zero lattice spacing.}.

We investigated the transition points for quark masses $m=0.1$ and $0.2$. 
The corresponding pseudoscalar and vector meson mass ratios 
are $m_{PS}/m_{V} \approx 0.70$ and $0.85$ \cite{Kar00}. 
We compute the Polyakov loop, chiral condensate, and their susceptibilities. 
The simulations were performed on a $16^3 \times 4$ lattice for
7 values of 
$\beta\in[3.64,3.67]$ for $m=0.1$ and 6 values of 
$\beta\in[3.74,3.80]$ for $m=0.2$, using the Hybrid R algorithm. 
We adopted a step size $\Delta \tau = 0.25 \times m$ and a
molecular dynamics trajectory length $\tau=0.5$. 
For each trajectory
10 sets of Z$_2$ noise vectors were used to calculate 
the reweighting factor and the derivatives of $\bar{\psi} \psi$ up to 
second order in $\mu$. 

For the calculation of mass reweighting surveyed in Sec.~\ref{sec:mass}, 
we took a total of 220600 trajectories at $m=0.1$ and 
155000 trajectories at $m=0.2$.
For the study with $\mu\not=0$ described 
in Sec.~\ref{sec:mu}, we used 128000 trajectories at $m=0.1$ and 
86000 trajectories at $m=0.2$.
The details are summarized in Table~\ref{tab:para}. 
The multi-histogram method of \cite{Swen88} was used to reweight 
in the $\beta$-direction using data from several values of $\beta$. 
Errors were estimated using 
the jack-knife method 
with bin size 100 trajectories. 

\section{Reweighting for quark mass}
\label{sec:mass}

\begin{figure}[t]
\centerline{
\epsfxsize=11.0cm\epsfbox{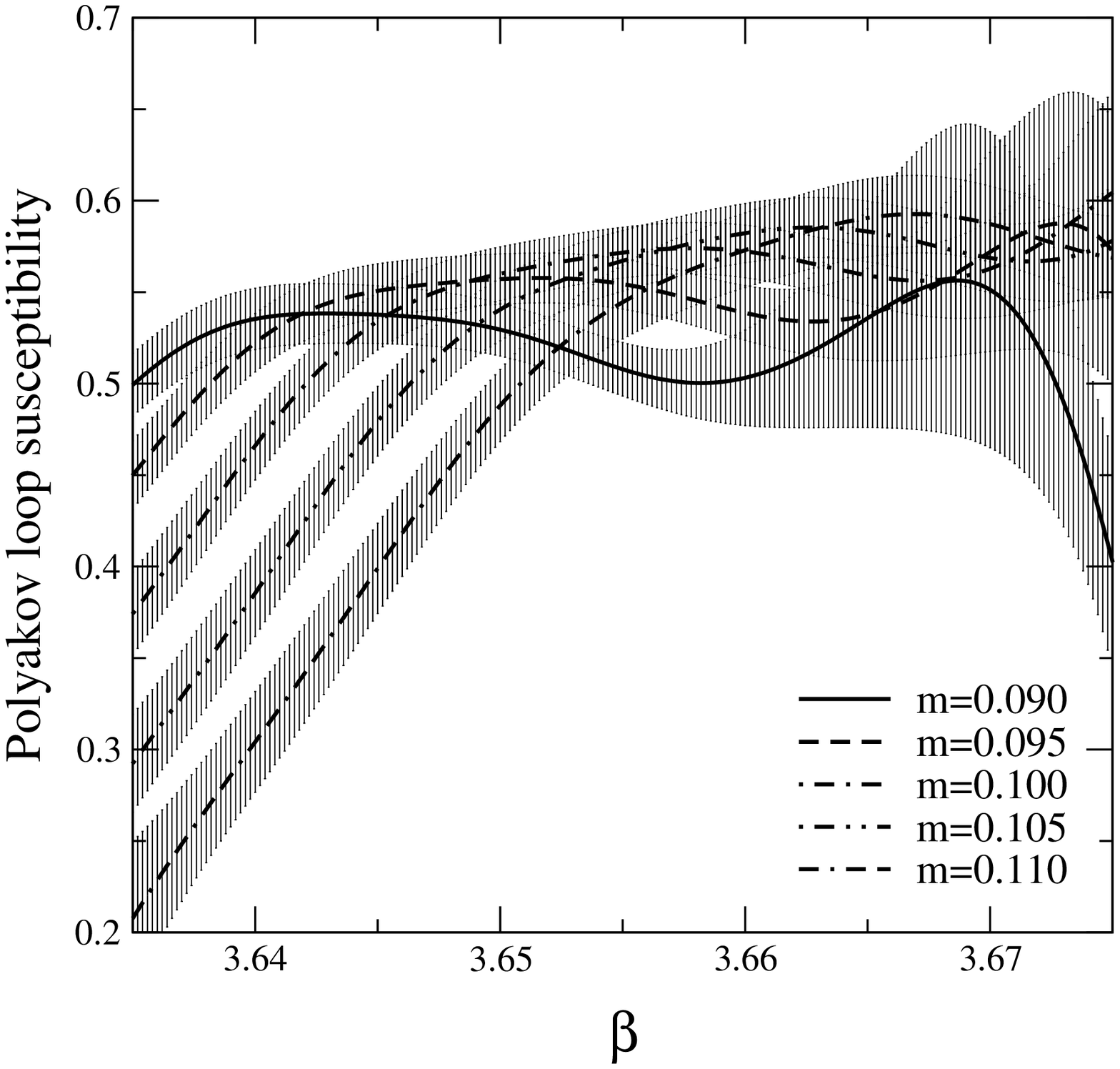}
}
\caption{
Quark mass dependence of $\chi_L$ 
as a function of $\beta$ at $m_0=0.1$.
}
\vspace*{-4mm}
\label{fig:psu01m}
\end{figure}

\begin{figure}[t]
\centerline{
\epsfxsize=11.0cm\epsfbox{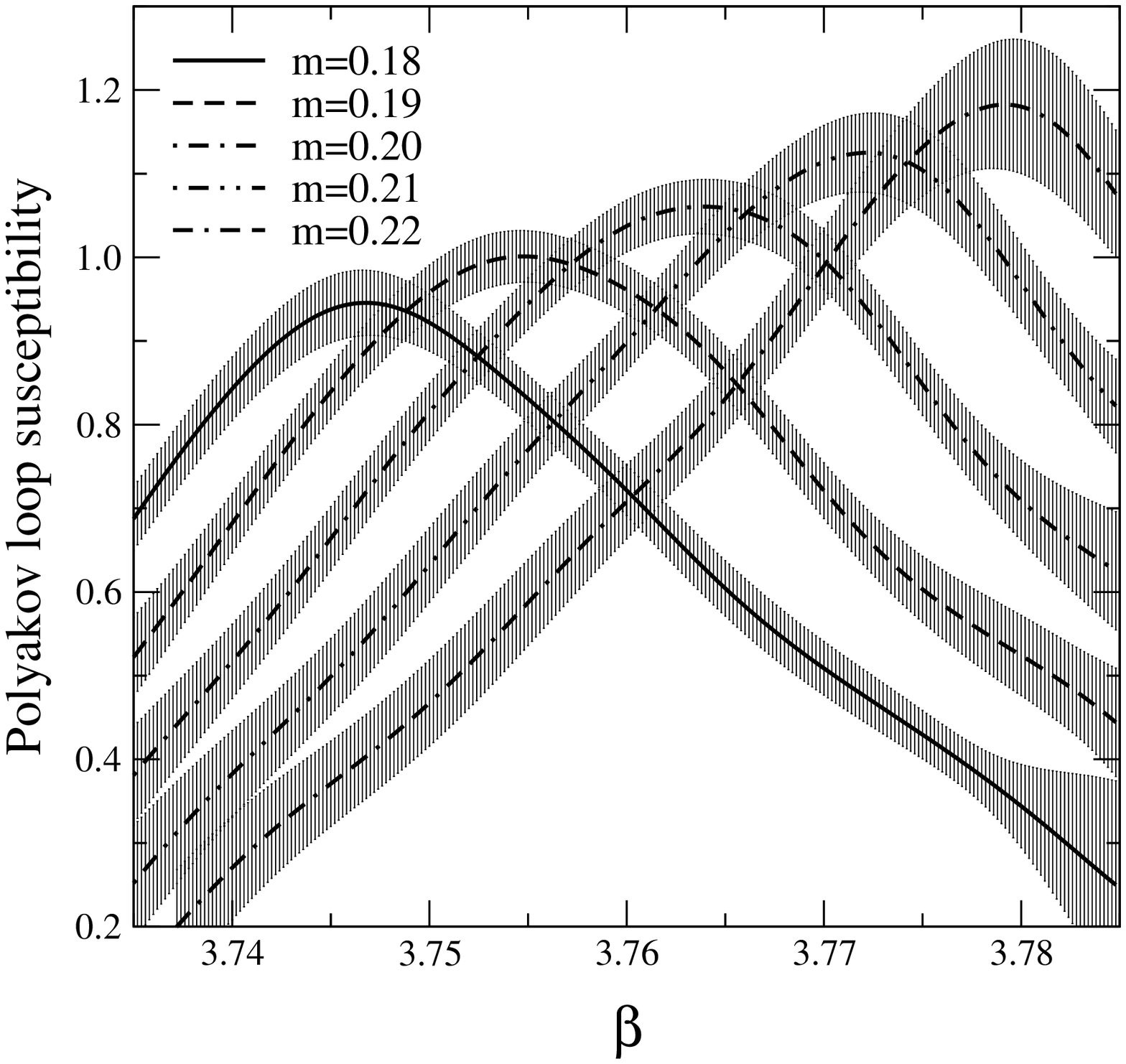}
}
\caption{
Quark mass dependence of $\chi_L$
as a function of $\beta$ at $m_0=0.2$.
}
\vspace*{-4mm}
\label{fig:psu02m}
\end{figure}

\begin{figure}[t]
\centerline{
\epsfxsize=11.0cm\epsfbox{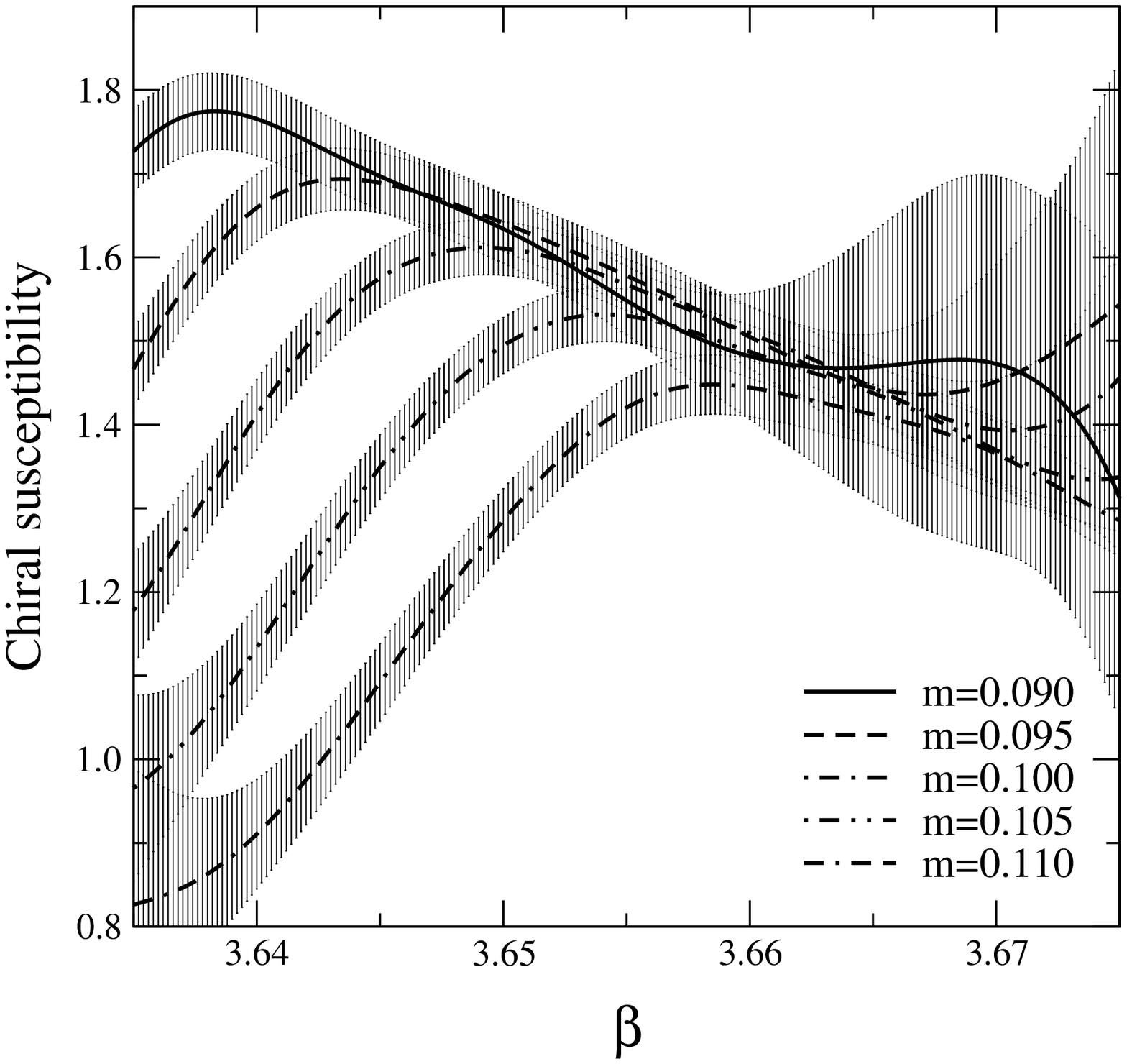}
}
\caption{
Quark mass dependence of $\chi_{\bar\psi\psi}$ as a function of $\beta$ 
at $m_0=0.1$.
}
\vspace*{-4mm}
\label{fig:csu01m}
\end{figure}

\begin{figure}[t]
\centerline{
\epsfxsize=11.0cm\epsfbox{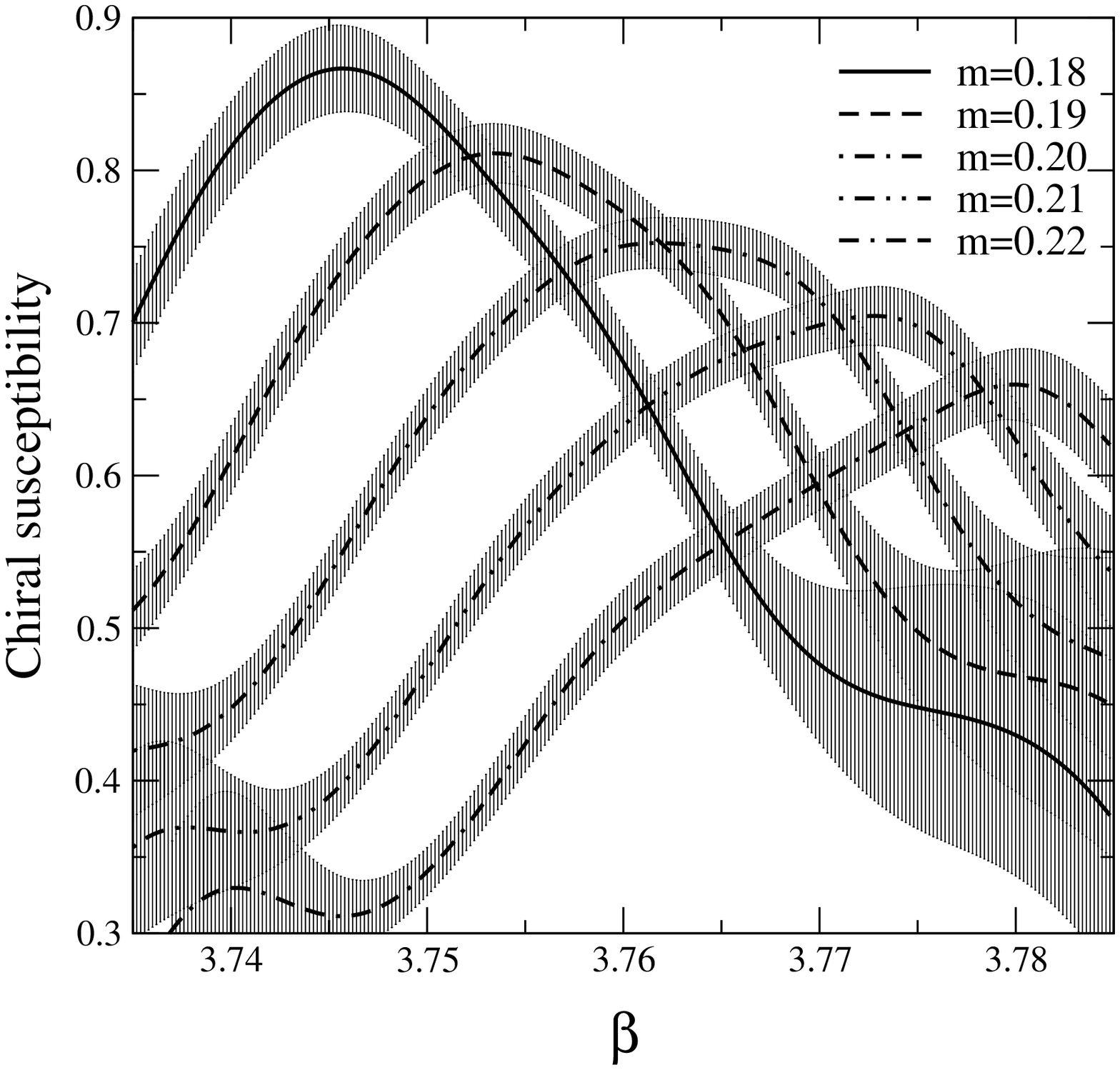}
}
\caption{
Quark mass dependence of $\chi_{\bar\psi\psi}$ as a function of $\beta$ 
at $m_0=0.2$.
}
\vspace*{-4mm}
\label{fig:csu02m}
\end{figure}

\begin{figure}[t]
\centerline{
\epsfxsize=11.0cm\epsfbox{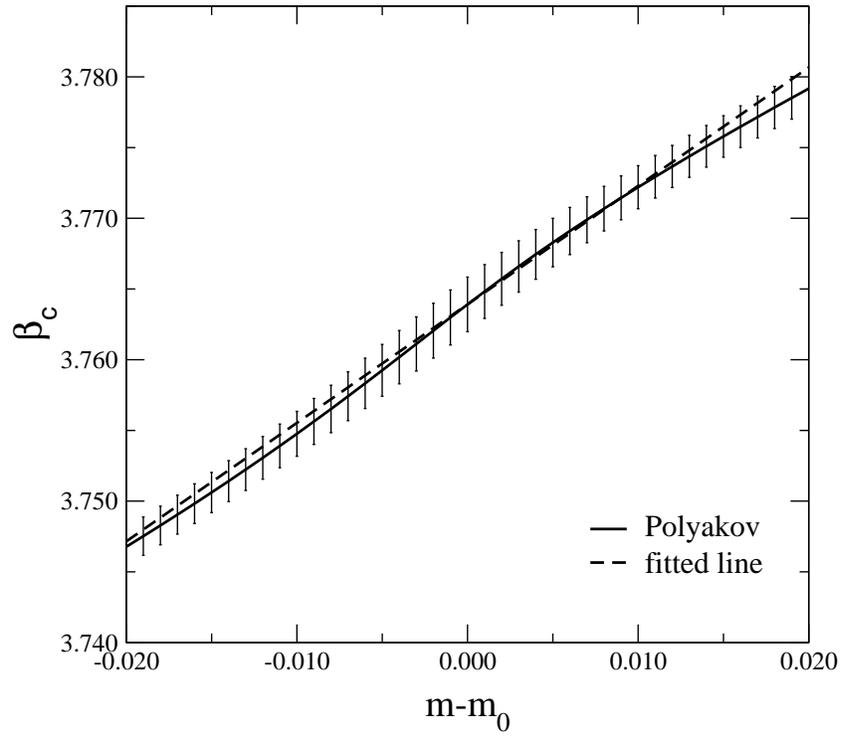}
}
\caption{
$\beta_c(m)$ determined by $\chi_L$
around $m_0=0.2$.
}
\vspace*{-4mm}
\label{fig:bcpm02}
\end{figure}

\begin{figure}[t]
\centerline{
\epsfxsize=11.0cm\epsfbox{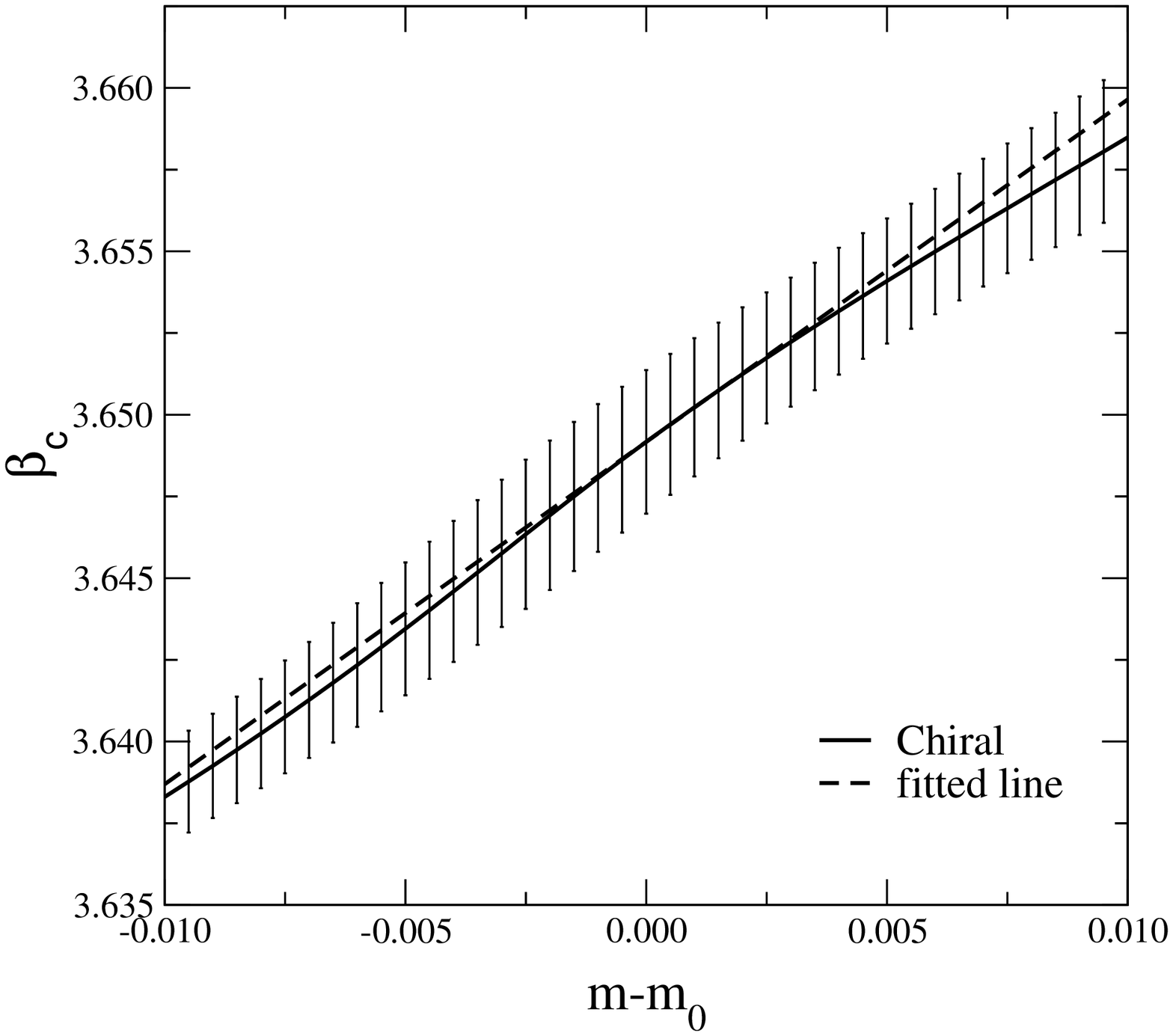}
}
\caption{
$\beta_c(m)$ determined by $\chi_{\bar\psi\psi}$
around $m_0=0.1$.
}
\vspace*{-4mm}
\label{fig:bccm01}
\end{figure}

\begin{figure}[t]
\centerline{
\epsfxsize=11.0cm\epsfbox{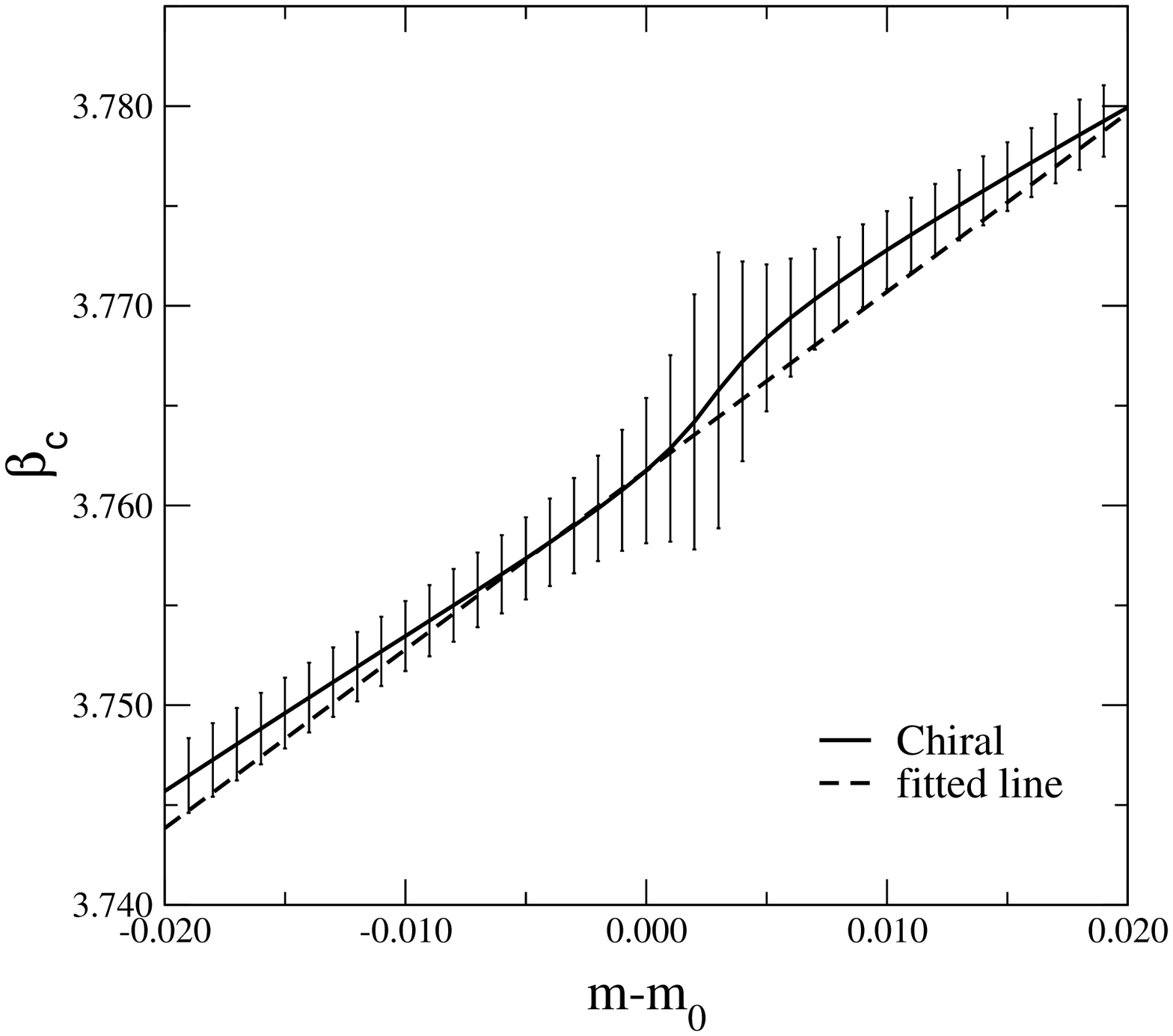}
}
\caption{
$\beta_c(m)$ determined by $\chi_{\bar\psi\psi}$
around $m_0=0.2$.
}
\vspace*{-4mm}
\label{fig:bccm02}
\end{figure}

\begin{figure}[t]
\centerline{
\epsfxsize=11.0cm\epsfbox{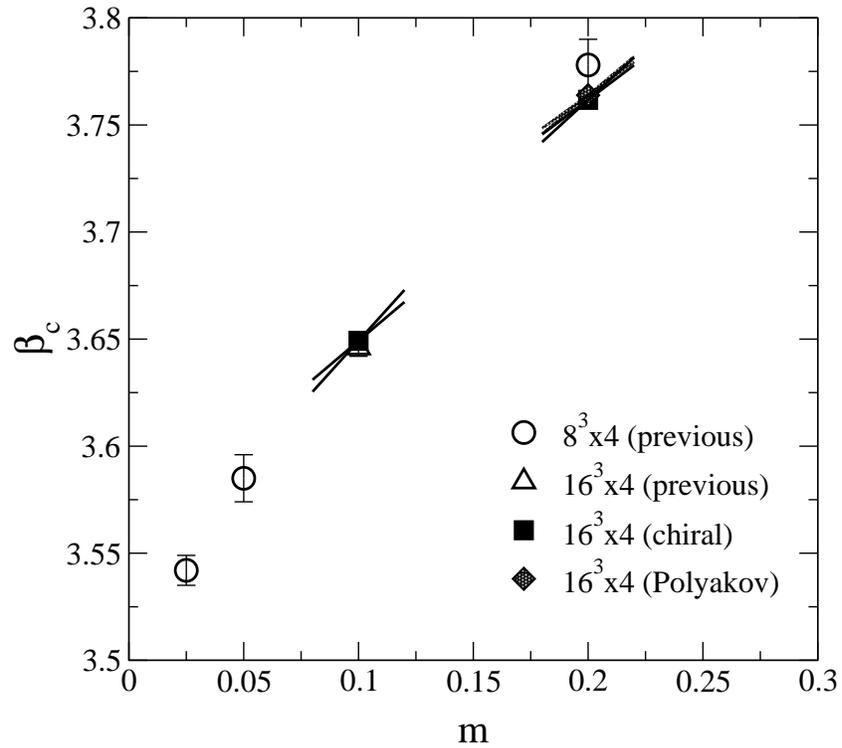}
}
\caption{
$\beta_c(m) $ determined by $\chi_{\bar\psi\psi}$
in comparison with previous results. 
}
\vspace*{-4mm}
\label{fig:bcall}
\end{figure}

Before calculating derivatives with respect to $\mu$, 
it is worthwhile to calculate the derivatives with respect to quark mass $m$, 
which is not only potentially 
important for the chiral extrapolation, but also a good 
demonstration of the reweighting technique for a parameter appearing
in the fermion action.
Because we cannot compare the result obtained by reweighting 
in the $\mu$ direction with the result of an actual simulation at $\mu\not=0$, 
this test is a necessary 
check of the reliability of our method.
The reweighting formula for quark mass is easily obtained 
from eqn.~(\ref{eq:expand}) and eqns.~(\ref{eq:dermu}) 
by replacing $\partial^n M/\partial \mu^n$ with 
$\partial M/\partial m=1$ and 
$\partial^n M/\partial m^n=0$ for $n \geq 2$. 
In the case of the reweighting for $m$,
we compute the fermionic reweighting factor up to second order, 
and the chiral condensate up to first order, i.e.
\begin{eqnarray}
\ln \det M(m) - \ln \det M(m_0) &=& {\rm tr} M^{-1}(m-m_0) - 
{\rm tr} (M^{-1} M^{-1}) (m-m_0)^2/2 +O[(m-m_0)^3],\\
\bar{\psi} \psi &=& (N_s^3 N_t)^{-1} \alpha N_{\rm f} [ {\rm tr} M^{-1} 
- {\rm tr} (M^{-1} M^{-1}) (m-m_0)] +O[(m-m_0)^2]. 
\end{eqnarray}
Hence, the error of the Polyakov loop susceptibility is $O[(m-m_0)^3]$ 
and that of the chiral susceptibility $O[(m-m_0)^2]$. 
Figures \ref{fig:psu01m} and \ref{fig:psu02m} show $\chi_L$
and Figs.~\ref{fig:csu01m} and \ref{fig:csu02m} show $\chi_{\bar\psi\psi}$
for different $m$ as functions of $\beta$ 
for simulation masses  $m_0=0.1$ and $0.2$. 
These figures show that the peak position moves to smaller $\beta$ as 
$m$ decreases as expected.
Moreover we find that as $m$ decreases 
the peak height becomes lower for $\chi_L$ 
and higher for $\chi_{\bar\psi\psi}$.
These behaviours are consistent since
the Polyakov loop is an exact 
order parameter only in the limit $m\to\infty$,
while the chiral condensate 
is an order parameter only for $m\to0$.
The phase transition is known to be a crossover for 
2 flavor QCD with $m>0$.
We calculate the slope of the transition point ${\partial\beta_c/\partial m}$
assuming that $\beta_c(m)$ is defined by the  peak 
position of the susceptibility whenever a clear peak is 
obtained\footnote{
Because the peak width of $\chi_L$ is too wide 
for the smaller mass $m=0.1$, we do not determine the pseudocritical
point for $L$ in this case.}.
Figures \ref{fig:bcpm02}, \ref{fig:bccm01} and \ref{fig:bccm02} 
show $\beta_c(m)$ for each $m_0$. 
We fit the data by a power series expansion about 
$m_0$, i.e., 
$\beta_c(m)=\beta_c(m_0) + \sum_{n=1}^{N_{\rm fit}} c_n (m-m_0)^n$,
with fit range $ |(m-m_0)/m_0| \leq 0.05$ or $0.1$. 
The results are presented in Table \ref{tab:mass}. 
We find a linear fit to be adequate with 
the dependence on choice of $N_{\rm fit}$ less than $3 \%$;
the discrepancy from the choice of fit range is less than $10 \%$. 
Both uncertainties lie well within the statistical error.
We denote the fitted line for $N_{\rm fit}=1$ and $ |(m-m_0)/m_0| \leq 0.1$ 
by a dashed line.  In Fig.~\ref{fig:bcall}
we compare the predicted variation of $\beta_c(m)$ with previously
existing data \cite{Kar00}.
Filled symbols are the results of the current study. The short lines denote 
the upper and lower bounds on the slope $\beta_c^\prime$.
From this figure, we find that reweighting 
yields results which are quite consistent with those of direct simulation, 
and hence infer that reweighting the fermion action 
using the technique we have outlined works well.

\section{Reweighting for chemical potential}
\label{sec:mu}

\subsection{Chemical potential dependence of the transition temperature}
\label{sec:bcmu}

\begin{figure}[t]
\centerline{
\epsfxsize=11.0cm\epsfbox{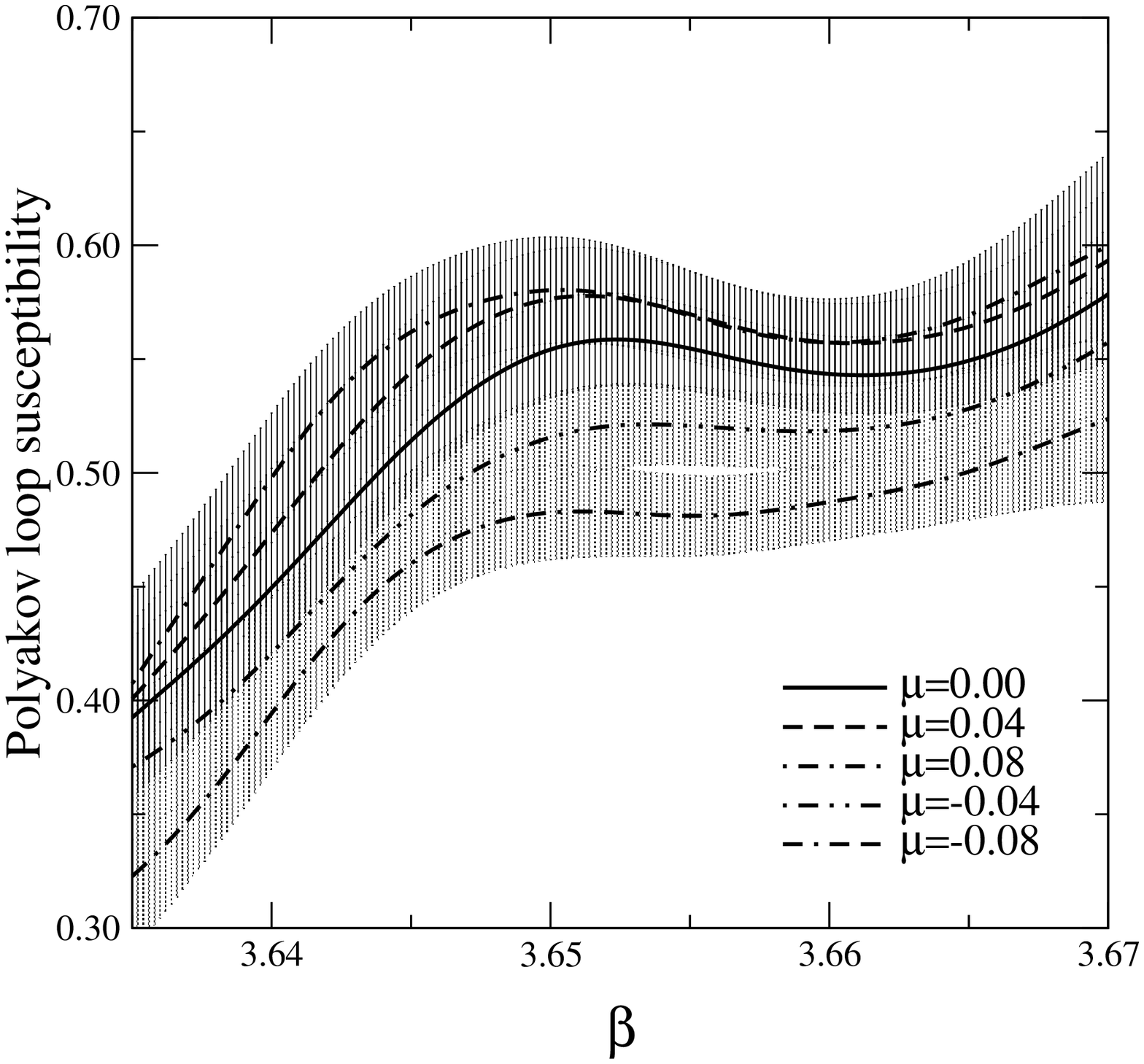}
}
\caption{
$\chi_L(\beta)$ at $m=0.1$ 
for various $\mu$.
}
\vspace*{-4mm}
\label{fig:psu01}
\end{figure}

\begin{figure}[t]
\centerline{
\epsfxsize=11.0cm\epsfbox{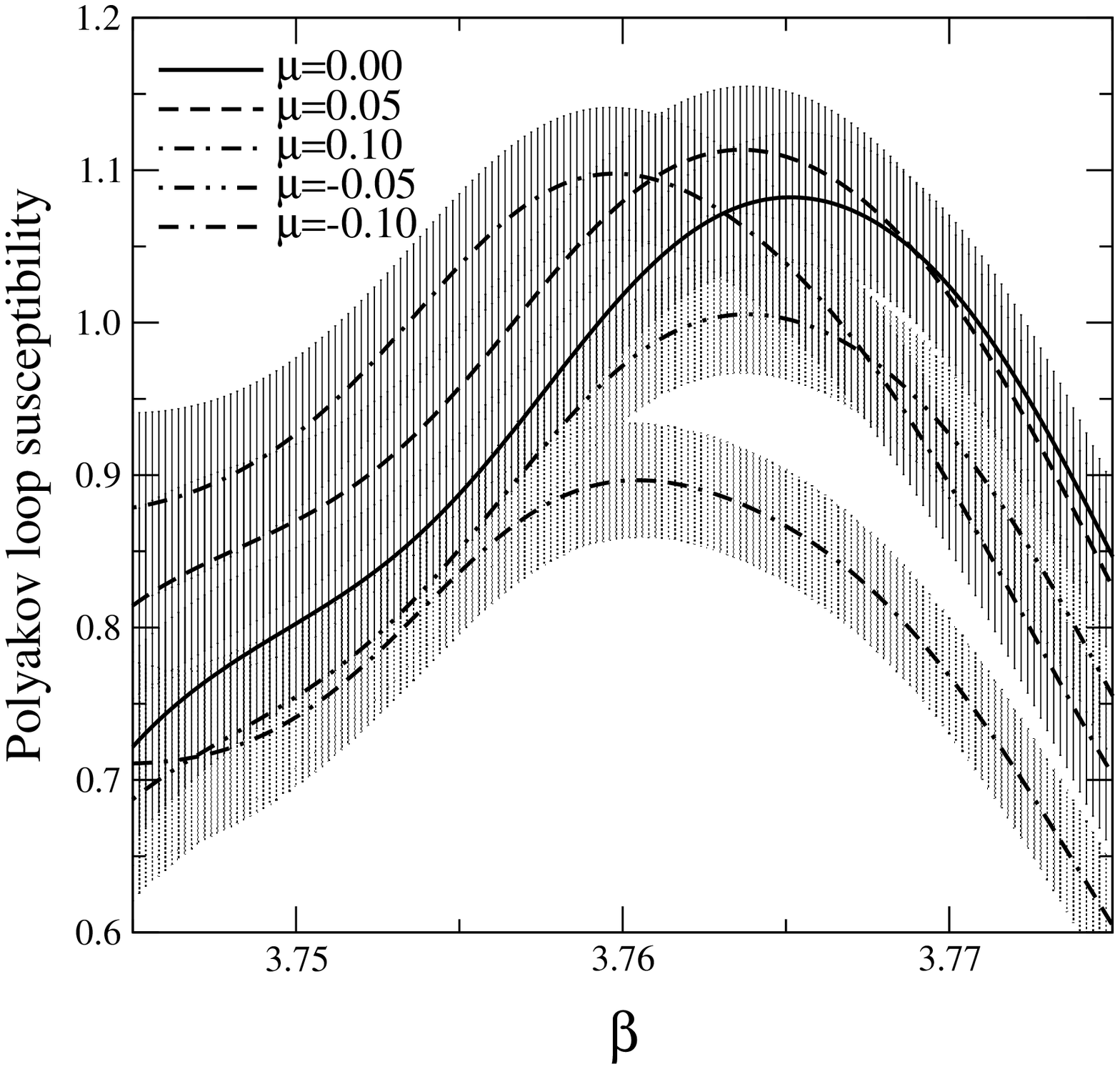}
}
\caption{
$\chi_L(\beta)$ at $m=0.2$ 
for various $\mu$.
}
\vspace*{-4mm}
\label{fig:psu02}
\end{figure}

\begin{figure}[t]
\centerline{
\epsfxsize=11.0cm\epsfbox{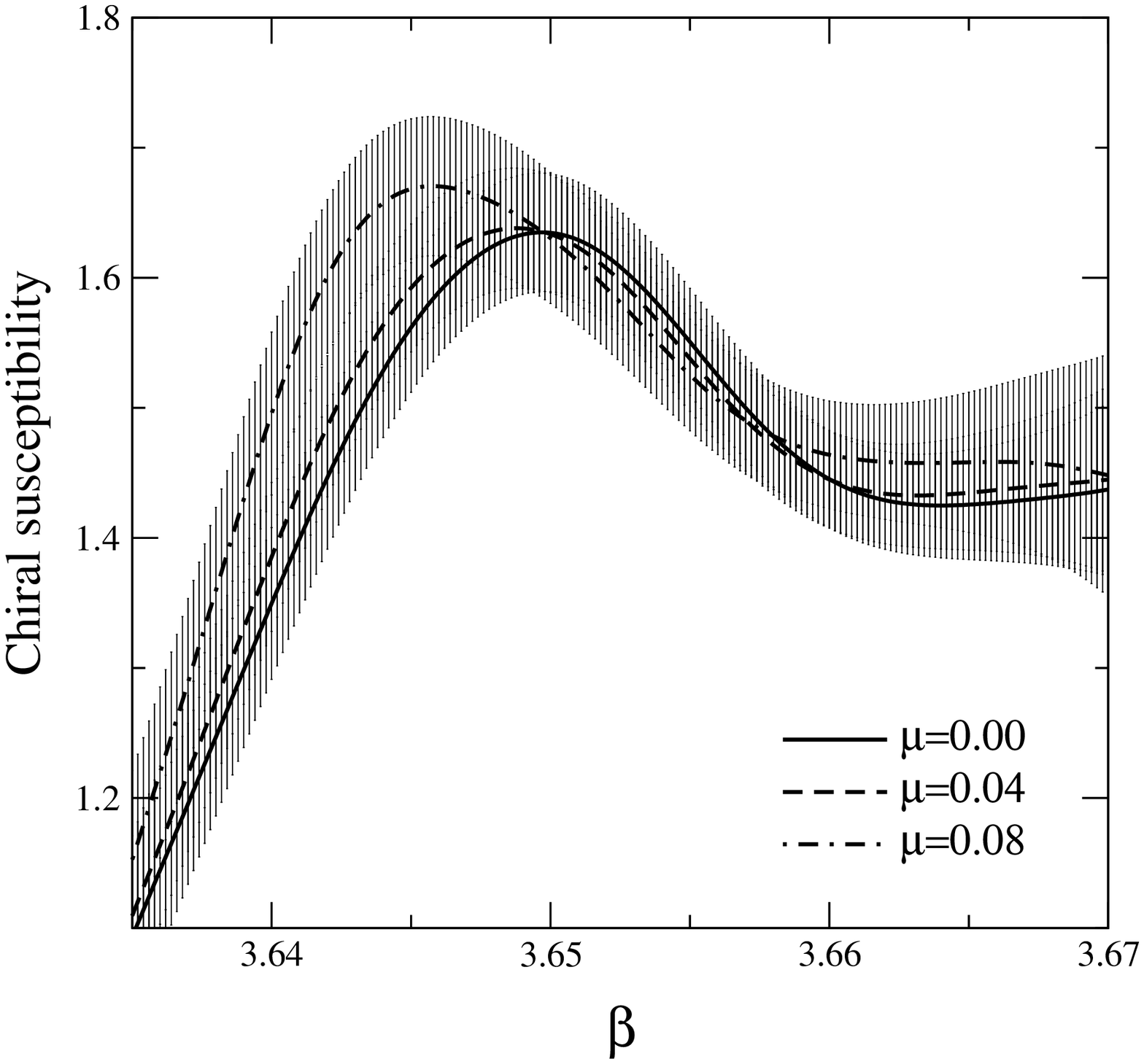}
}
\caption{
$\chi_{\bar\psi\psi}(\beta)$ at $m=0.1$ for various $\mu$.
}
\label{fig:csu01}
\end{figure}

\begin{figure}[t]
\centerline{
\epsfxsize=11.0cm\epsfbox{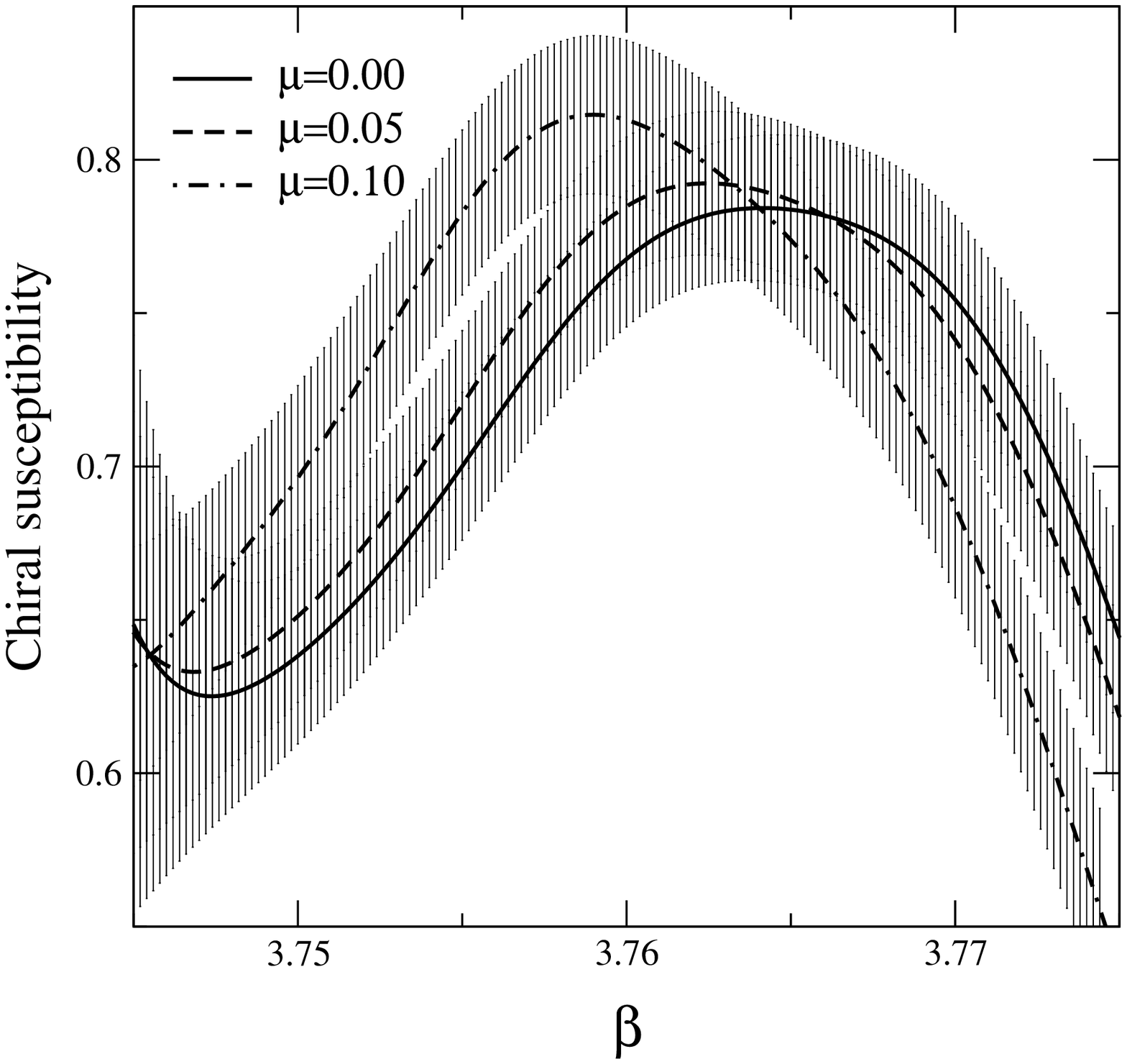}
}
\caption{
$\chi_{\bar\psi\psi}(\beta)$ at $m=0.2$ for various $\mu$.
}
\label{fig:csu02}
\end{figure}

\begin{figure}[t]
\centerline{
\epsfxsize=11.0cm\epsfbox{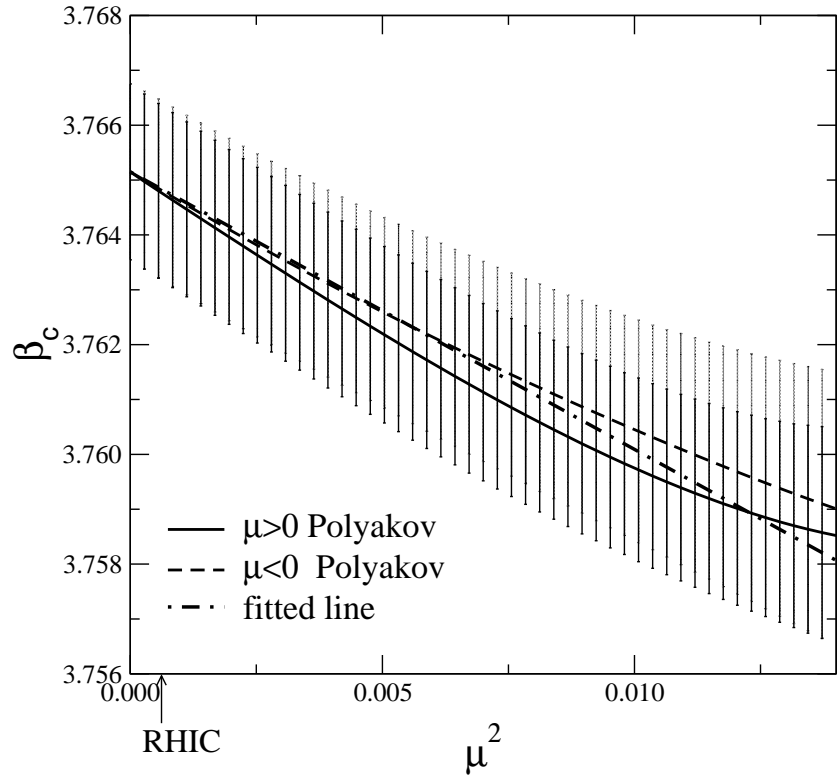}
}
\caption{
Phase transition point $\beta_c(\mu)$ determined by 
$\chi_L$ at $m=0.2$. 
}
\vspace*{-4mm}
\label{fig:bcp02}
\end{figure}

\begin{figure}[t]
\centerline{
\epsfxsize=11.0cm\epsfbox{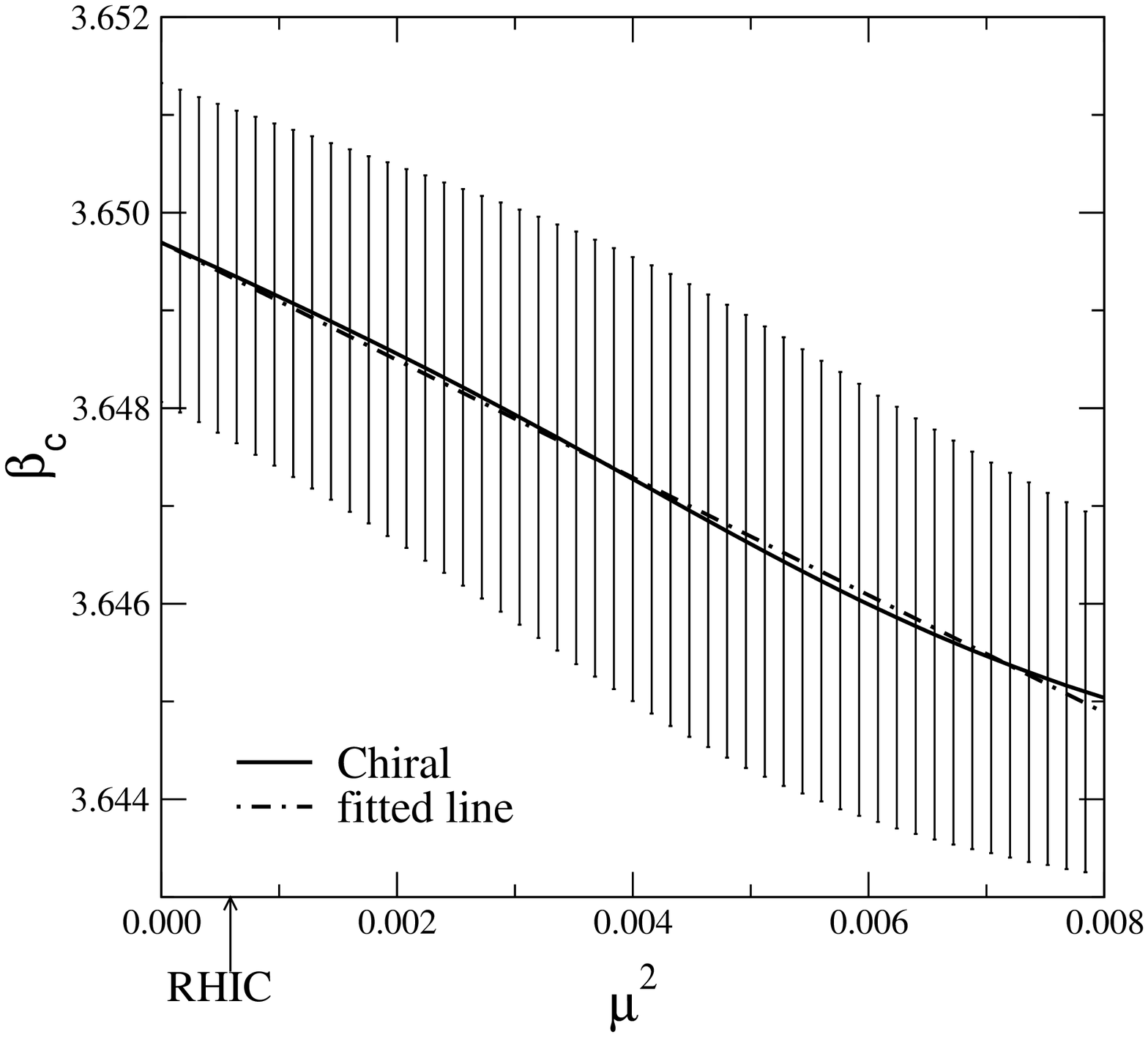}
}
\caption{
Phase transition point $\beta_c(\mu)$ determined by 
$\chi_{\bar\psi\psi}$ at $m=0.1$. 
}
\vspace*{-4mm}
\label{fig:bcc01}
\end{figure}

\begin{figure}[t]
\centerline{
\epsfxsize=11.0cm\epsfbox{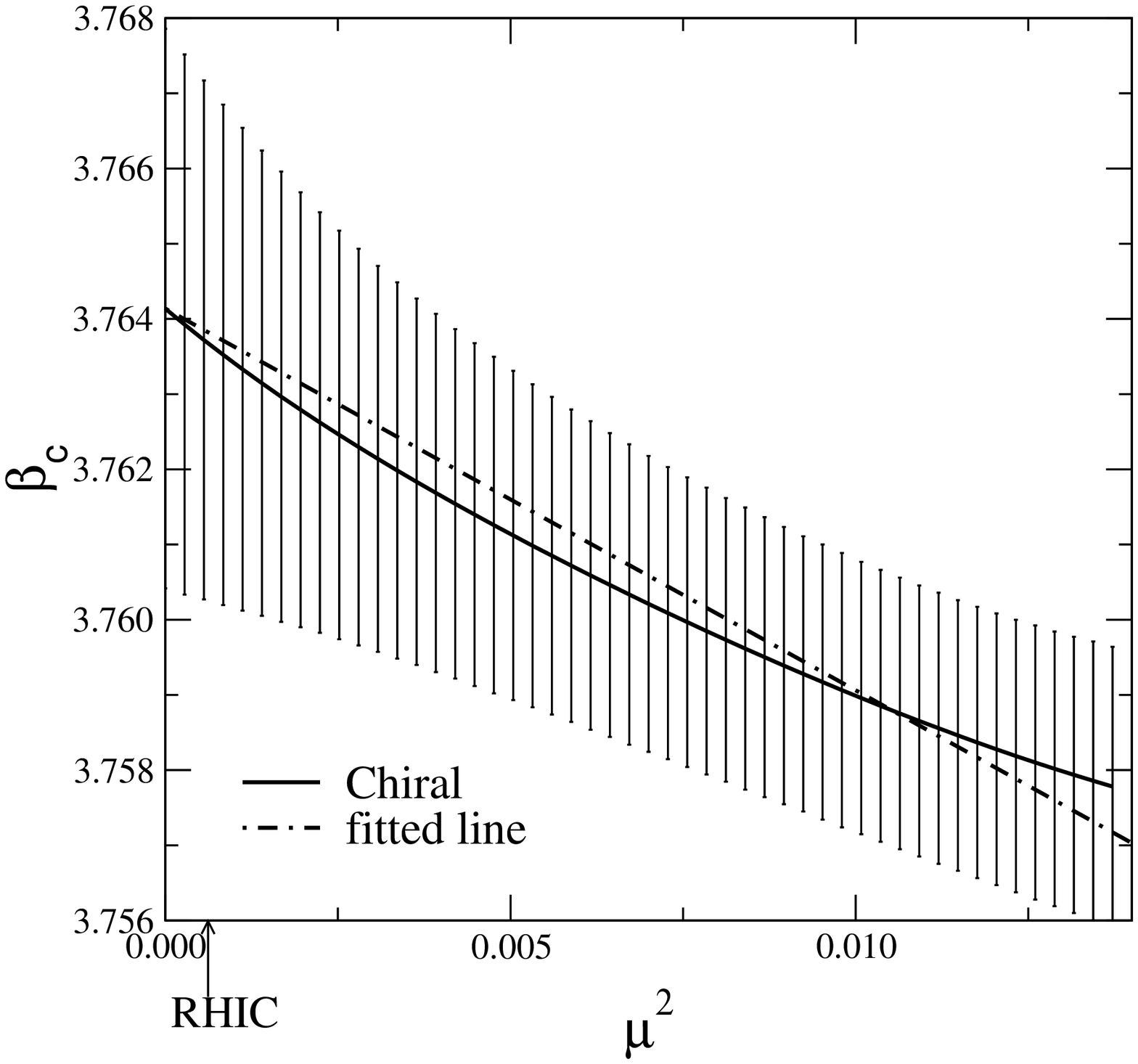}
}
\caption{
Phase transition point $\beta_c(\mu)$ determined by 
$\chi_{\bar\psi\psi}$ at $m=0.2$. 
}
\vspace*{-4mm}
\label{fig:bcc02}
\end{figure}

Next we turn our attention to reweighting with respect to $\mu$, with the 
Taylor expansion made about the simulation point $\mu=0$. 
First we calculate the derivatives of the transition point with respect 
to $\mu$ in the region of small $\mu$ relevant to RHIC. 
In Figs.~\ref{fig:psu01}, \ref{fig:psu02}, \ref{fig:csu01} and 
\ref{fig:csu02}, we plot 
$\chi_L$ and $\chi_{\bar\psi\psi}$ 
at $m=0.1$ and $0.2$ for various $\mu$.
As outlined in section~\ref{sec:method},
we compute consistently up to $O(\mu^2)$ and 
expect the results to contain errors at $O(\mu^4)$.
Strictly speaking, the $O(\mu^3)$ term does not vanish for $L$ 
since it is complex (see Sec.~\ref{sec:method}). 
However, we expect that $\chi_L$ and $\chi_{\bar\psi\psi}$
yield the same $\beta_c$ (see below) with error $O(\mu^4)$. 
The figures show that the position of the susceptibility peak moves 
lower as $\mu$ increases, which means that the critical temperature 
becomes lower as $\mu$ increases. 
As we obtained well-localised  peaks for $\chi_L$ at $m=0.2$ 
and $\chi_{\bar\psi\psi}$ at $m=0.1$ and $0.2$, 
we use these peak positions to determine the transition point $\beta_c$
as functions of $\mu^2$ in Figs.~\ref{fig:bcp02}, \ref{fig:bcc01}, 
and \ref{fig:bcc02}. 
Note that because the Polyakov loop is 
interpreted as an external quark current 
running in the positive time direction, 
positive and negative $\mu$ give different contributions to 
both $L$ and $\chi_L$, and we display both cases. 
Figs.~\ref{fig:bcp02}, \ref{fig:bcc01} and \ref{fig:bcc02} 
also display the value of $\mu=0.1T_c$ relevant for RHIC.
The shift $\beta_c(\mu)-\beta_c(0)$ is found to be small at this point.

Because the first derivative is expected to be zero as discussed above, 
we fit the $\beta_c$ data by a straight line in $\mu^2$, fixing 
$\beta_c$ at $\mu=0$, in ranges $\mu^2 \leq 0.008(0.014)$ for 
$m=0.1(0.2)$ respectively, in which the phase problem is not 
serious (see Sec.~\ref{sec:phase} below).
We obtain ${\rm d}^2 \beta_c/{\rm d} \mu^2 = -1.20(44)$ and $-1.02(56)$ 
at $m=0.1$ and $0.2$ from the chiral susceptibility and 
${\rm d}^2 \beta_c/{\rm d} \mu^2 = -1.01(23)$ at $m=0.2$ from 
the Polyakov loop susceptibility. 
Dot-dashed lines in Figs.~\ref{fig:bcp02}, \ref{fig:bcc01} and \ref{fig:bcc02} 
are the fitted lines. 
To investigate the fit range dependence and the fitting function dependence, 
we also tried the range $\mu^2 \leq 0.005(0.01)$ for $m=0.1(0.2)$,
and using a quadratic fit function.
Table~\ref{tab:betac} summarises the results.
We may conclude that 
$\vert{\rm d}^2 \beta_c/{\rm d} \mu^2\vert\approx 1.1$ with $30 - 50 \%$ error, 
and any quark mass dependence of ${\rm d}^2 \beta_c/{\rm d} \mu^2$ is not 
visible within the accuracy of our calculation. 

\begin{figure}[t]
\centerline{
\epsfxsize=11.0cm\epsfbox{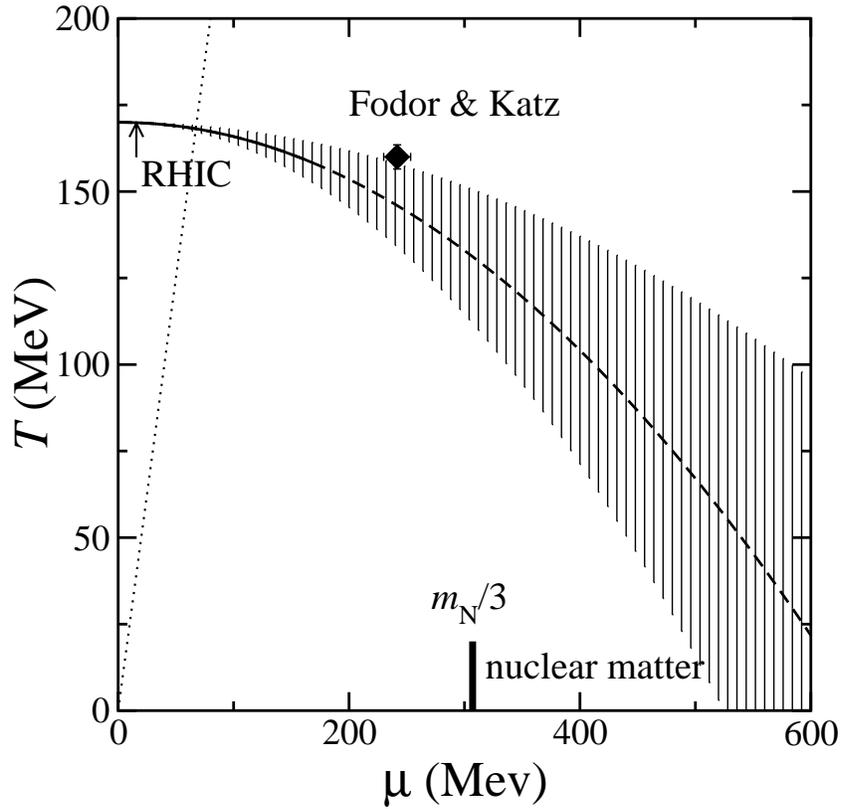}
}
\caption{
Sketch of the phase diagram, as estimated using our value of the curvature
of $\beta_c(\mu=0)$. 
The errors shown are statistical only and reflect
the uncertainty of the coefficient of the $O(\mu^2)$ term
in the expansion of $T_c(\mu)$.
Dotted line is $\mu / T =0.4$. 
The diamond symbol is the end point of the first order phase transition 
obtained by Fodor and Katz \protect\cite{Fod01}.
}
\vspace*{-4mm}
\label{fig:cur}
\end{figure}

Of course, it is desirable to translate these observations into physical units.
The second derivative of $T_c$ can be estimated by 
\begin{eqnarray}
\frac{{\rm d}^2 T_c}{{\rm d} \mu_{\rm q}^2} = -\frac{1}{N_t^2 T_c} 
\left. \frac{{\rm d}^2 \beta_c}{{\rm d} \mu^2} \right/ 
\left( a \frac{{\rm d} \beta}{{\rm d} a} \right),
\end{eqnarray}
where $a$ is the lattice spacing.
The beta-function may be obtained from the string tension data in 
Ref.\cite{Kar00}. 
We compute it by differentiating the interpolation function of the string 
tension 
with an ansatz\cite{allton96}:
\begin{eqnarray}
\sqrt{\sigma a^2} (\beta) = R(\beta)(1 + c_2 \hat{a}^2 (\beta) 
+ c_4 \hat{a}^4 (\beta)) / c_0,
\end{eqnarray}
where $R(\beta)$ is the usual 2-loop scaling function, 
$\hat{a} \equiv R(\beta)/R(\bar{\beta})$ and $\bar{\beta}=3.70$. 
$c_0, c_2$ and $c_4$ are fit parameters with 
$c_0=0.0570(35), c_2=0.669(208)$ and $ c_4=-0.0822(1088)$ 
at $m=0.1$. We get 
$a^{-1}({\rm d}a/{\rm d}\beta)=-2.08(43)$ at $(\beta,m)=(3.65,0.1)$. 
We then find 
$T_c({\rm d}^2 T_c/{\rm d}\mu^2_{\rm q}) \approx -0.14$ at $m=0.1$. 
We sketch the phase transition line with $50 \%$ error in Fig.~\ref{fig:cur}
assuming $T_c \simeq 170{\rm MeV}$.
In the figure we also indicate the line $\mu_{\rm q}/T=0.4$, corresponding
roughly to the range over which the fits to the leading order behaviour 
of $T_c(\mu)$ shown in Figs.~\ref{fig:bcp02} --~\ref{fig:bcc02} are made.
Of course, one has to expect that higher order terms in the
expansion become relevant for $\mu / T = O(1)$. To
quantify this we will have to analyze higher order contributions
in the expansion in the future. To indicate the present
systematic uncertainty in the transition line for larger $\mu / T$
we show this region as a dotted line in Fig.~\ref{fig:cur}.
We stress that the errors shown are statistical only and reflect
the uncertainty of the coefficient of the $O(\mu^2)$ term
in the expansion of $T_c(\mu)$.
On the assumption that the transition line is parabolic
all the way down to $T=0$, then
this curvature is too small to be consistent with
the phenomenological 
expectation that at $T=0$ a transition from hadronic to quark matter occurs 
for $\mu_c$ some 50 - 200 MeV greater than the onset of nuclear matter
at $\mu_o\simeq m_N/3\simeq307$MeV \cite{phen}.
This tendency is also supported by the result of Fodor and Katz 
\cite{Fod01}, and hints at contributions from
higher-order derivatives, or even non-analytic behaviour, at larger values of
$\mu$.
Despite the large errors 
we can see that our result gives us useful information about 
the phase diagram, at least for small $\mu$, 
because the first derivative is zero. 

\begin{figure}[t]
\centerline{
\epsfxsize=11.0cm\epsfbox{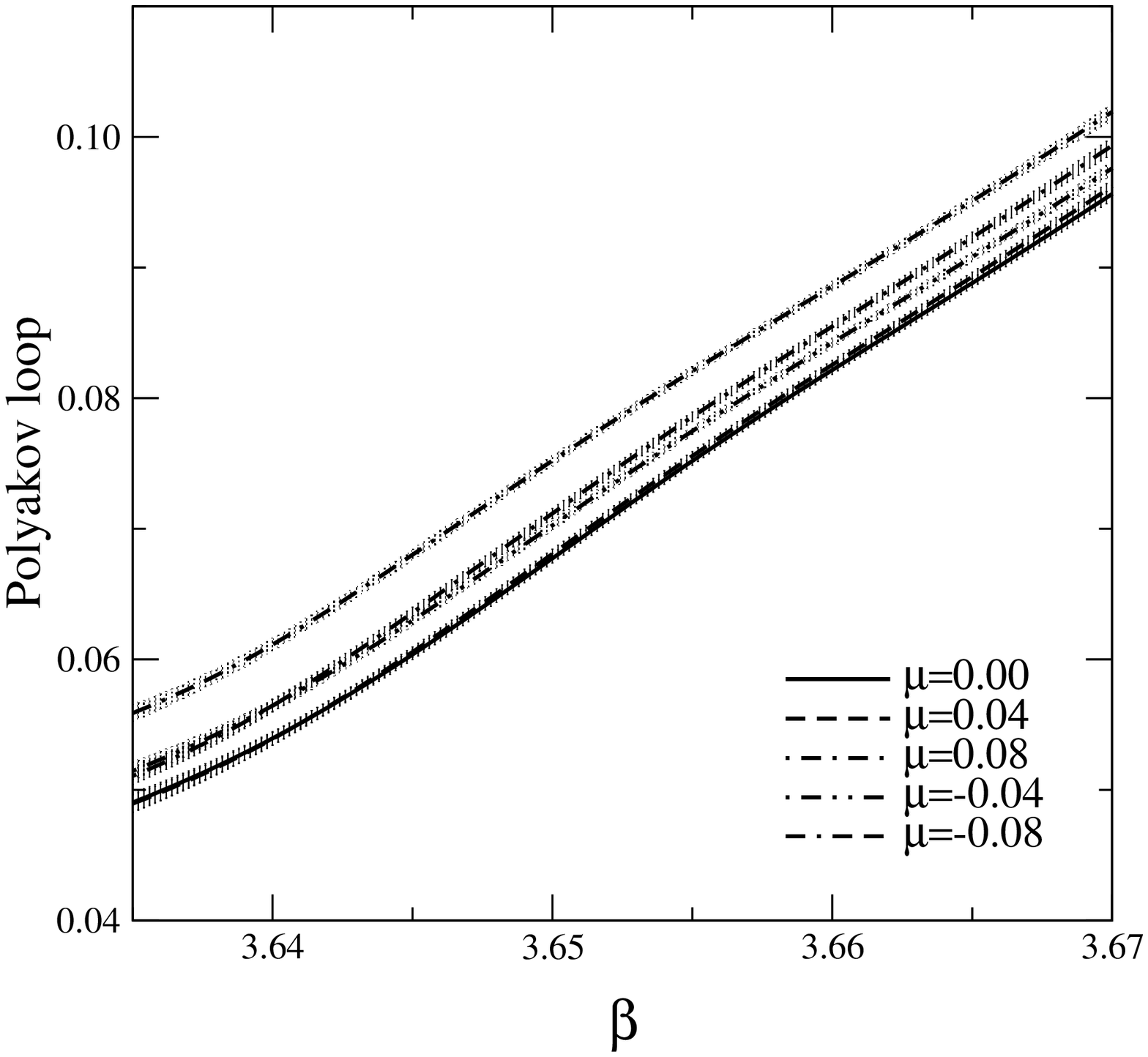}
}
\caption{
$L(\beta)$ at $m=0.1$ 
for various $\mu$.
}
\vspace*{-4mm}
\label{fig:pol01}
\end{figure}

\begin{figure}[t]
\centerline{
\epsfxsize=11.0cm\epsfbox{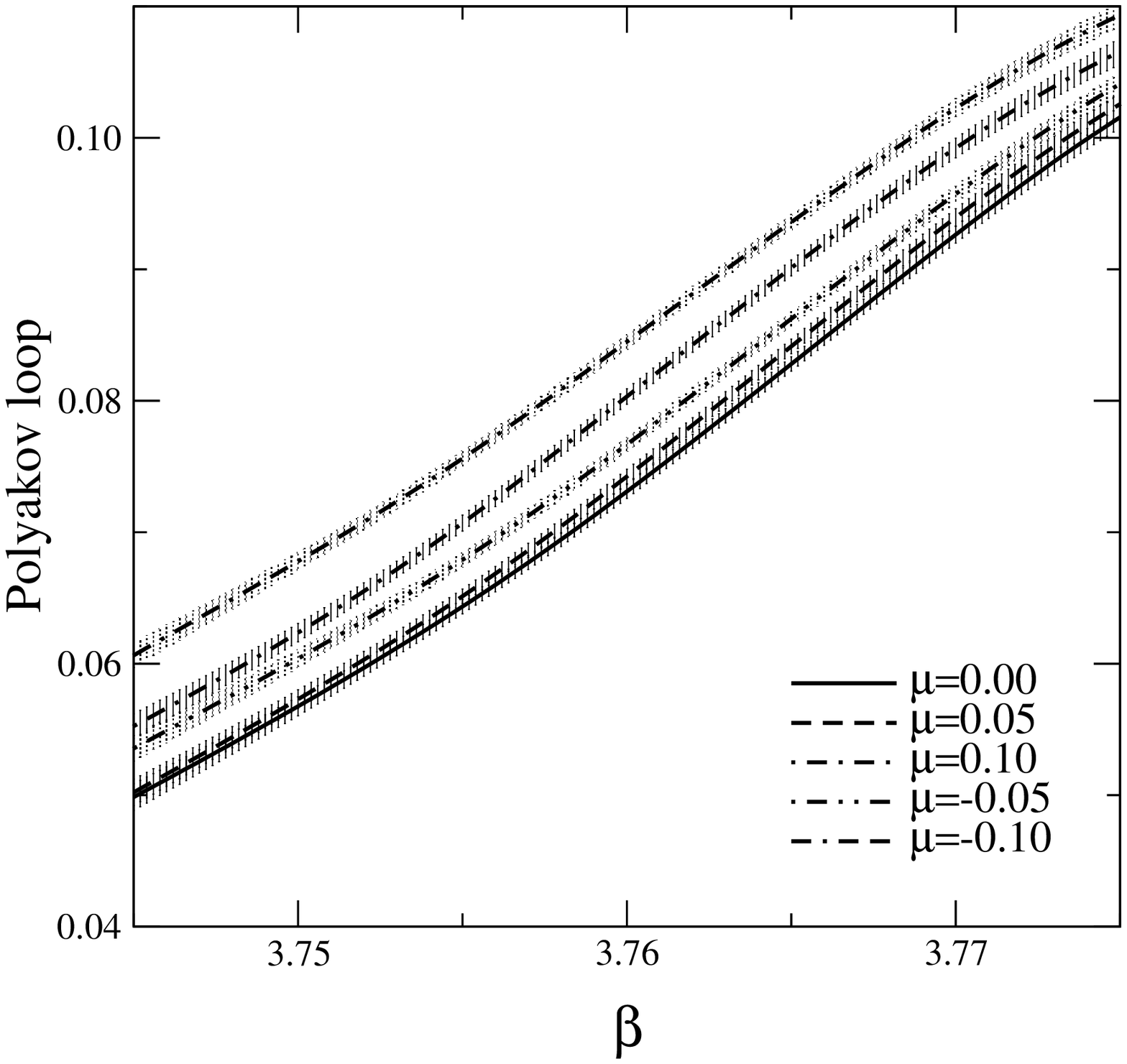}
}
\caption{
$L(\beta)$ at $m=0.2$ 
for various $\mu$.
}
\vspace*{-4mm}
\label{fig:pol02}
\end{figure}

Another point worth noting is the screening effect of dynamical anti-quarks 
at $\mu < 0$. 
Negative chemical potential induces the dynamical generation 
of anti-quarks, which in contrast to quarks can completely 
screen an external color triplet current. 
Thus the free energy of a single quark is reduced, especially 
in the confinement phase, and the singularity at the phase transition point 
is weakened due to the reduction in the range of current-current interactions. 
This effect can be seen in Figs.~\ref{fig:psu01}, \ref{fig:psu02}, 
\ref{fig:pol01} and \ref{fig:pol02}, where 
we denote the Polyakov loop and its susceptibility at $\mu<0$ by 
dot-dot-dash and dot-dash-dash lines. 
We see that $L$ at $\mu<0$ is larger than that 
at $\mu>0$, which means that the free energy at $\mu<0$ 
is smaller. Moreover, as seen in Fig.~\ref{fig:psu02}, the peak height of 
$\chi_L$ becomes smaller for $\mu<0$, while
the position of the pseudocritical line in Fig.~\ref{fig:bcp02} is almost 
the same between positive and negative $\mu$. 
The screening effect only seems to make the phase transition singularity 
weaker without shifting the pseudocritical line. 
Because the only source of asymmetry between $\mu$ and $-\mu$ is due to
the correlation between the imaginary parts of the fermion determinant 
and $L$, these imaginary contributions help to decrease the susceptibility 
at $\mu < 0$. In this way, we can see that the explicit breaking of time
reversal symmetry by exchange of $\mu$ with $-\mu$ helps to highlight
an interesting feature of dynamical quarks 
in full QCD.

\begin{figure}[t]
\centerline{
\epsfxsize=11.0cm\epsfbox{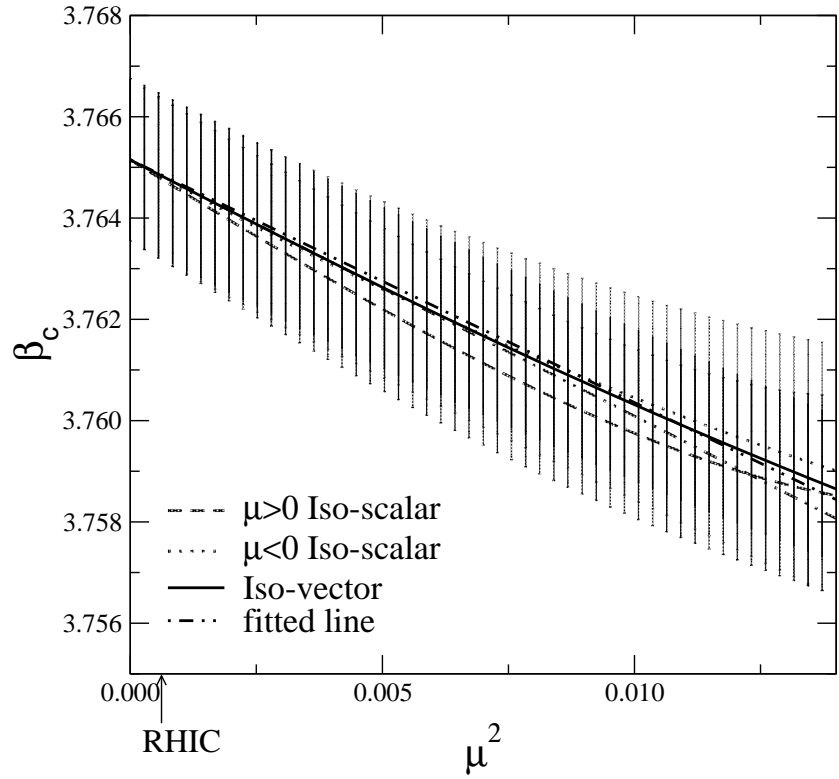}
}
\caption{
Difference between $\mu$ and $\mu_I$ 
for $\beta_c$ determined by $\chi_L$
at $m=0.2$. 
}
\vspace*{-4mm}
\label{fig:ivbcp02}
\end{figure}

\begin{figure}[t]
\centerline{
\epsfxsize=11.0cm\epsfbox{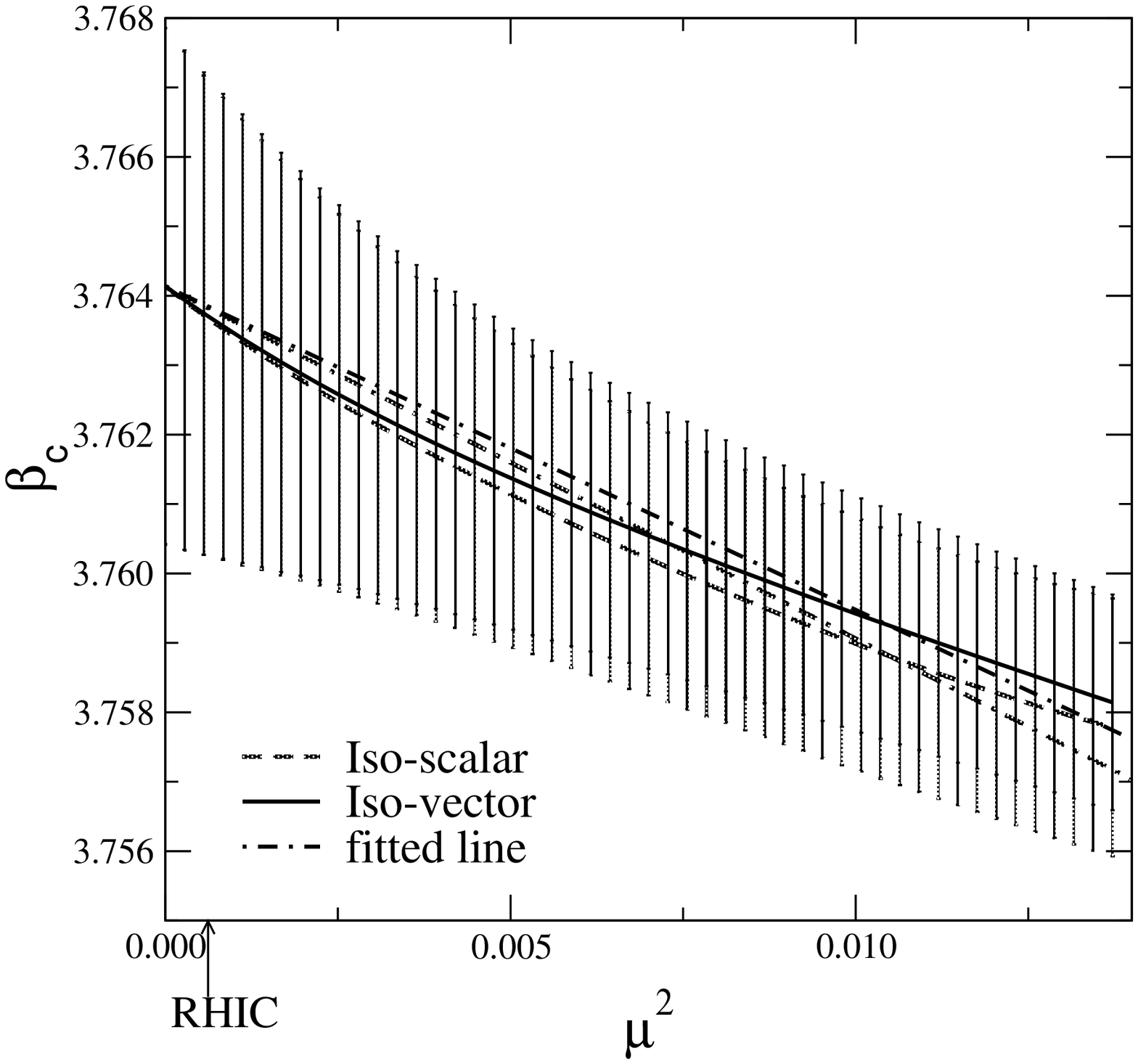}
}
\caption{
Difference between $\mu$ and $\mu_I$ 
for $\beta_c$ determined by $\chi_{\bar\psi\psi}$
at $m=0.2$. 
}
\vspace*{-4mm}
\label{fig:ivbcc02}
\end{figure}

\begin{figure}[t]
\centerline{
\epsfxsize=11.0cm\epsfbox{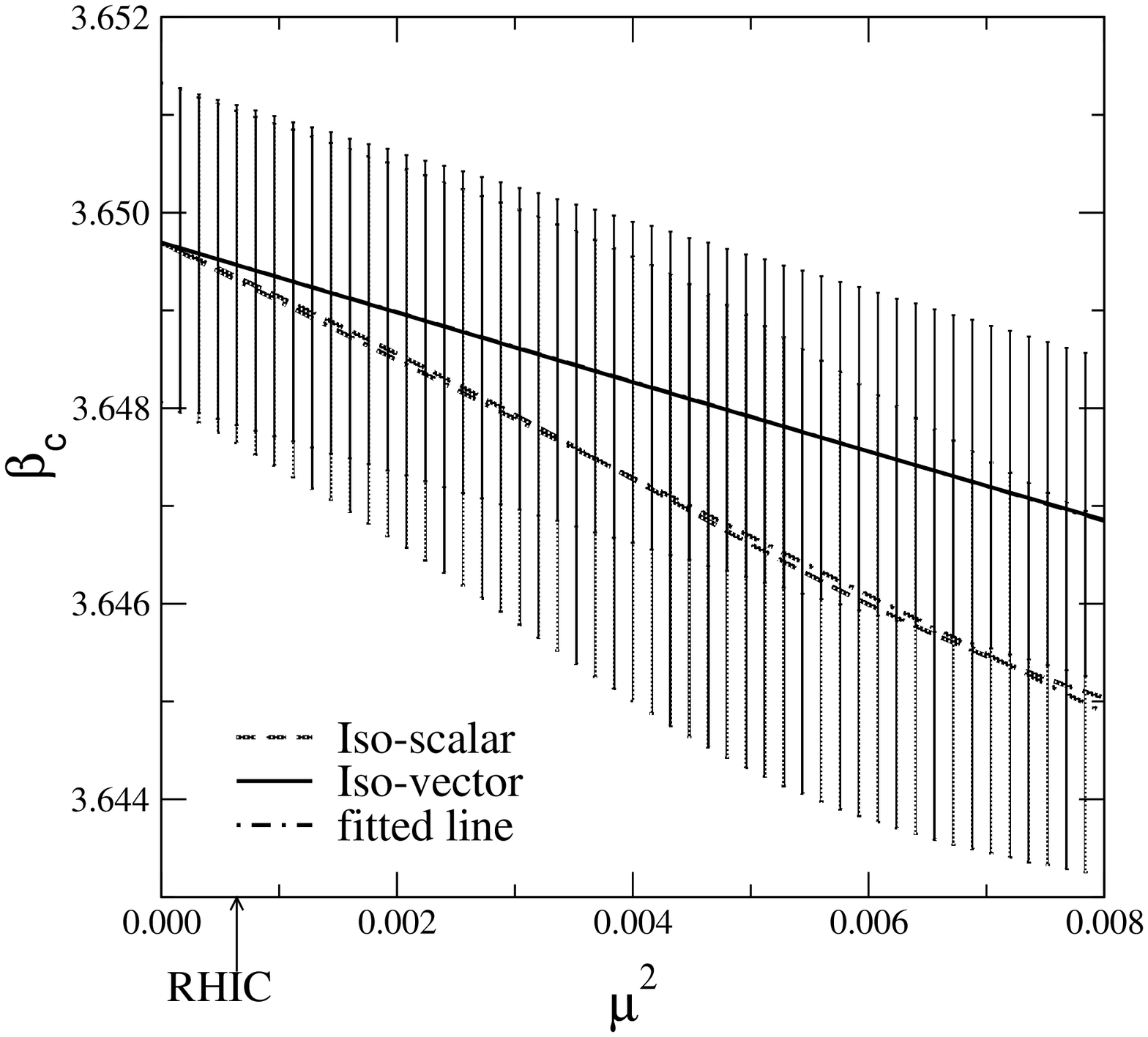}
}
\caption{
Difference between $\mu$ and $\mu_I$ 
for $\beta_c$ determined by $\chi_{\bar\psi\psi}$
at $m=0.1$. 
}
\vspace*{-4mm}
\label{fig:ivbcc01}
\end{figure}

Finally, if instead we were to 
impose an {\sl isovector\/} chemical potential $\mu_I$ 
having opposite sign for $u$ and 
$d$ quarks \cite{Gav01,SS}, 
then the quark determinant would become real and positive, enabling
simulations using standard Monte-Carlo methods \cite{KogSin}.
This motivates a comparison between systems with the usual isoscalar
chemical potential $\mu$ and the isovector chemical potential $\mu_I$.
In the framework of the Taylor expansion, terms even in $\mu$ are identical
for both $u$ and $d$ quarks, but odd terms 
cancel for the case $\mu_I\not=0$, meaning that terms proportional to
${\cal O}_1,{\cal R}_1$ should be set to zero in eqn.~(\ref{eq:specific}).
We analyzed the transition point $\beta_c(\mu_I)$ for $m=0.2$; 
the results are shown in Fig.~\ref{fig:ivbcp02} 
for $\beta_c$ determined by $\chi_L$
and Fig.~\ref{fig:ivbcc02} for that by $\chi_{\bar\psi\psi}$.
The solid line shows $\beta_c$ as a function of $\mu_I$, the 
dashed line $\beta_c(\mu)$. 
The second derivative of $\beta_c$ with respect to $\mu_I$ is found to be 
$-0.96(19)$ for $\chi_L$ and $-0.93(52)$ for $\chi_{\bar\psi\psi}$.
Dot-dashed lines in Figs~\ref{fig:ivbcp02} and \ref{fig:ivbcc02} 
show the fits.
Within errors 
there appears to be no significant difference 
between isovector and isoscalar chemical potentials
for small $\mu$. 
A similar analysis for $\chi_{\bar\psi\psi}$ 
at $m=0.1$ is shown in Fig.~\ref{fig:ivbcc01}; 
here the second derivative of $\beta_c$ is 
$-0.71(16)$, which is smaller than the isoscalar case.
However this result is also smaller than that obtained 
at $m=0.2$, which is physically unacceptable since the second 
derivative should approach zero as $m \to \infty$. 
Hence the difference between $\mu_I$ and $\mu$
at $m=0.1$ is most likely due to statistical error.

The terms we have dropped
are associated with fluctuations in the phase of $\det M$, which are
small in the region of small $\mu$, as will be demonstrated
in Sec.~\ref{sec:phase} below.
This is perhaps not unexpected on physical grounds --  increasing $\mu_I$
is predicted to induce the onset of matter in the form of a
pion condensate at a critical $\mu_{Io}\simeq m_{PS}/2$ \cite{SS}, 
and indeed evidence for this scenario in the form of a negative curvature 
for $m_{PS} (\mu_I)$ in the low $T$ phase is reported in \cite{Taro01}. 
However, even for $m=0.1$ on this lattice this scale is roughly 
$0.92 \sqrt{\sigma} \simeq 390{\rm MeV}$ \cite{Kar00}, which is 
a little larger than the isoscalar onset threshold $\mu_o\simlt m_N/3$.
The curvature with repect to $\mu_I$ should dominate 
as the chiral limit is approached and pion and nucleon mass scales become
separate. If this turns out to be the case, then it is interesting 
to note that phase correlations between observable and measure
actually {\em decrease} the physical effect of raising $\mu$;
this has also been observed in simulations of two-color QCD with
a single flavor of staggered adjoint quark \cite{TCQCD}, in which including
the sign of the fermion determinant has the effect of postponing
the onset transition.

\subsection{Quark number susceptibility and equation of state at $\mu\not=0$}
\label{sec:quarknum}

\begin{figure}[t]
\centerline{
\epsfxsize=11.0cm\epsfbox{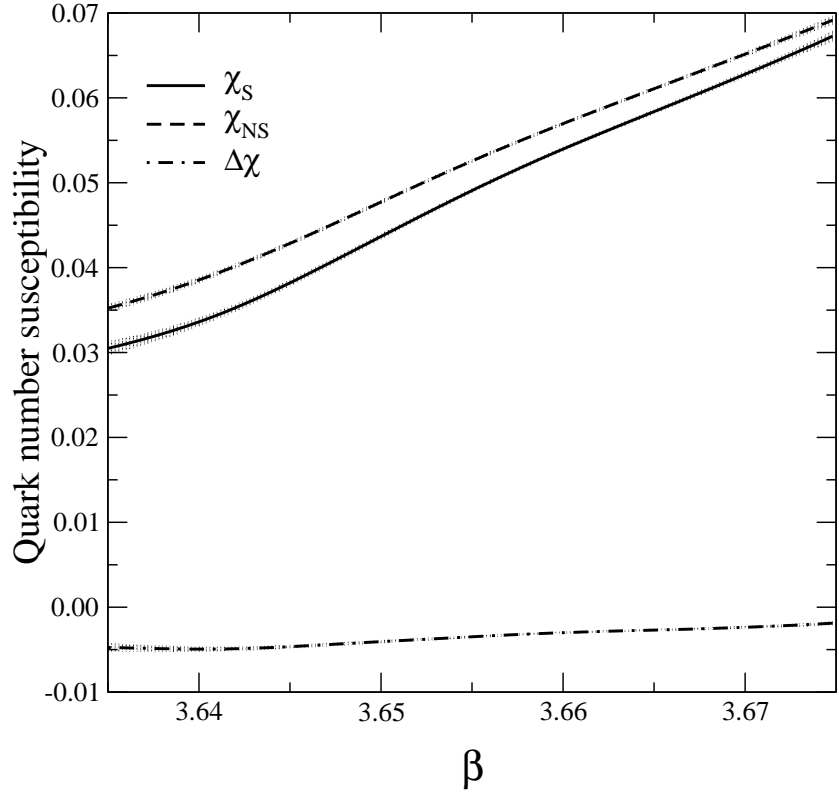}
}
\caption{
Quark number susceptibilities $\chi_{\rm S}$ and $\chi_{\rm NS}$ at $m=0.1$. 
$\Delta \chi = \chi_{\rm S} - \chi_{\rm NS}$.
}
\vspace*{-4mm}
\label{fig:qns01}
\end{figure}

\begin{figure}[t]
\centerline{
\epsfxsize=11.0cm\epsfbox{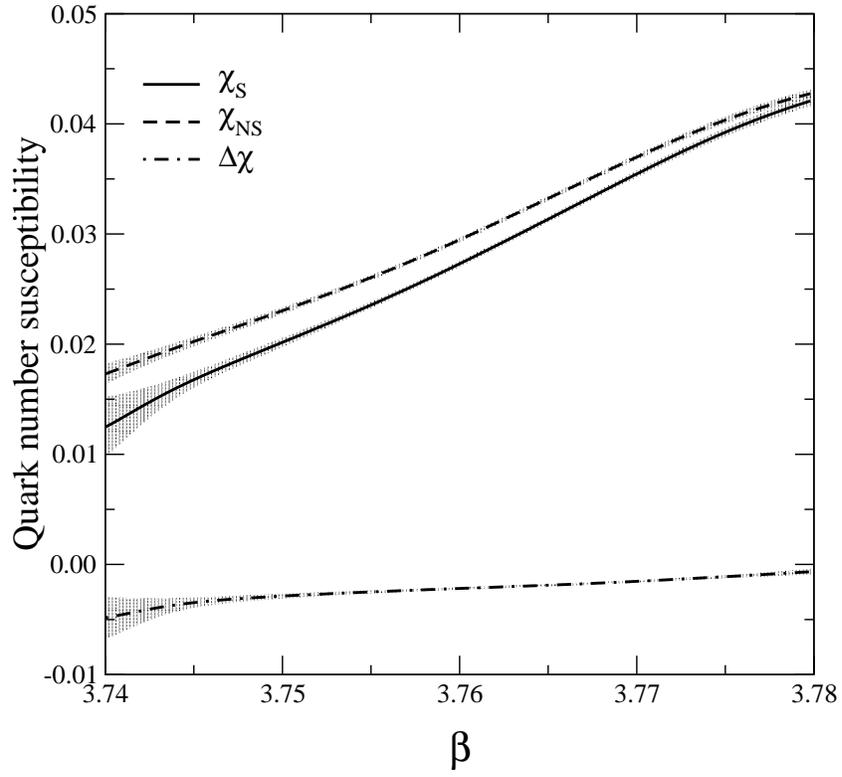}
}
\caption{
Quark number susceptibilities $\chi_{\rm S}$ and $\chi_{\rm NS}$ at $m=0.2$. 
}
\vspace*{-4mm}
\label{fig:qns02}
\end{figure}

The energy density $\epsilon$ and pressure $p$ at the critical point are 
interesting quantities for heavy-ion collision experiments. 
In this section, we discuss the $\mu$-dependence of 
the equation of state which describes them. 
If we employ the integral method based on the 
homogeneity of the system \cite{integral}, we obtain 
$p=(T/V)\ln{\cal Z}$;
derivatives of $p$ with respect to $\mu$ are then related to
the quark number density $n_{\rm q}$ (via a Maxwell relation)
and the singlet quark number susceptibility
$\chi_{\rm S}=\partial n_{\rm q}/\partial\mu_{\rm q}$
\cite{Gott88}: 
\begin{eqnarray}
\label{eq:eosder1}
\frac{\partial (p/T^4)}{\partial \mu_{\rm q}} 
&=& 
\frac{1}{VT^3} \frac{\partial \ln {\cal Z}}{\partial \mu_{\rm q}} 
= \frac{n_{\rm q}}{T^4} \\
\label{eq:eosder2}
\frac{\partial^2 (p/T^4)}{\partial \mu_{\rm q}^2} 
&=&
\frac{1}{VT^3} \frac{\partial^2 \ln {\cal Z}}{\partial \mu_{\rm
q}^2} 
= \frac{\chi_{\rm S}}{T^4}
\end{eqnarray}
Here $n_{\rm q}$, 
$\chi_{\rm S}$ and also the nonsinglet susceptibility $\chi_{\rm NS}$ are 
given in physical units by
\begin{eqnarray}
\label{eq:qnd}
{n_{\rm q}\over T^3} &=& 
\frac{\alpha N_{\rm f}N_t^2}{N_s^3} \left\langle {\rm tr} \left(M^{-1} 
\frac{\partial M}{\partial \mu} \right) \right\rangle. \\
{\chi_{\rm S}\over T^2}
&=& \frac{\alpha N_{\rm f}N_t}{N_s^3} \left[ \left\langle {\rm tr} 
\left(M^{-1} \frac{\partial^2 M}{\partial \mu^2} \right) \right\rangle 
- \left\langle {\rm tr} \left(M^{-1} \frac{\partial M}{\partial \mu} 
M^{-1} \frac{\partial M}{\partial \mu} \right) \right\rangle \right] 
\nonumber \\
&& + \frac{(\alpha N_{\rm f})^2N_t}{N_s^3} \left[ \left\langle {\rm tr} 
\left(M^{-1} \frac{\partial M}{\partial \mu} \right) {\rm tr} \left(M^{-1} 
\frac{\partial M}{\partial \mu} \right) \right\rangle 
- \left\langle {\rm tr} \left(M^{-1} \frac{\partial M}{\partial \mu} 
\right) \right\rangle^2 \right], \\
{\chi_{\rm NS}\over T^2}
&=& \frac{\alpha N_{\rm f}N_t}{N_s^3} \left[ \left\langle {\rm tr} 
\left(M^{-1} \frac{\partial^2 M}{\partial \mu^2} \right) \right\rangle 
- \left\langle {\rm tr} \left(M^{-1} \frac{\partial M}{\partial \mu} 
M^{-1} \frac{\partial M}{\partial \mu} \right) \right\rangle \right],
\end{eqnarray}
The quark number density is zero at $\mu=0$ so once again the leading correction
is $O(\mu^2)$.
The susceptibilities $\chi_{\rm S} a^2$, 
and $\chi_{\rm NS} a^2$ are plotted in Figs. \ref{fig:qns01} and 
\ref{fig:qns02} for $m=0.1$ and $0.2$. 
Because $\chi_{\rm S} a^2 = 0.0433(3)$ and $0.0306(2)$ for $m=0.1$ and 
$0.2$ at $\beta_c$ in Table~\ref{tab:betac} $(\bar{\psi}\psi)$, we obtain 
$T^2 \partial^2(p/T^4)/\partial \mu_{\rm q}^2 = 0.693(5)$ $(m=0.1)$ and 
0.490(4) $(m=0.2)$ at $\beta_c$.
The discrepancy of $p/T^4$ at the interesting point for RHIC, 
$\mu_{\rm q}/T_c \sim 0.1$, from its value at $\mu=0$ 
is about $0.0035(0.0024)$ for $m=0.1(0.2)$;
since $p/T^4 \approx 0.27$ 
at $\beta_c$ for  $(m,\mu) = (0.1,0)$ \cite{Kar00} this is a 1\% effect, and
hence quite small. We can also obtain estimates of the quark number density
via $n_q a^3\simeq\mu_{\rm q}a\chi_Sa^2$, with results
$n_{\rm q}/T^3=0.693(5)\mu_{\rm q}/T$ and $0.490(4)\mu_{\rm q}/T$ 
for $m=0.1$ and $0.2$ which 
assuming $T\simeq170$MeV translates into roughly 9\% and 6\% 
of nuclear matter density at the RHIC point. Clearly these values will need 
careful extrapolation to the
chiral limit before a meaningful comparison with experiment can be made.

Moreover the energy density $\epsilon$ can also be estimated via the equation
for the conformal anomaly:
\begin{eqnarray}
\label{eq:e-3p}
\frac{\epsilon - 3p}{T^4} 
&=& -\frac{1}{VT^3} a \frac{\partial \ln {\cal Z}}{\partial a} \nonumber \\
&=& -\frac{1}{VT^3} \left[ a \frac{\partial \beta}{\partial a} 
\frac{\partial \ln {\cal Z}}{\partial \beta} +
a \frac{\partial m}{\partial a} \frac{\partial \ln {\cal Z}}{\partial m} 
\right].
\end{eqnarray}
Here we estimate $\epsilon$ in the chiral limit, where
$a\partial m/\partial a$ can be neglected. We find 
\begin{eqnarray}
\label{eq:e-3pap}
\frac{\epsilon - 3p}{T^4} 
\approx -\frac{1}{VT^3} \frac{\partial \ln {\cal Z}}{\partial \beta} 
\left( \frac{1}{a} \frac{\partial a}{\partial \beta} \right)^{-1},
\end{eqnarray}
with second derivative
\begin{eqnarray}
\label{eq:e-3pd2}
\frac{\partial^2 ((\epsilon -3p)/T^4)}{\partial \mu_{\rm q}^2} 
\approx - \frac{1}{T^4} \frac{\partial \chi_{\rm S}}{\partial \beta}
 \left( \frac{1}{a} \frac{\partial a}{\partial \beta} \right)^{-1}.
\end{eqnarray}
Because the quark mass dependence of the equation of state seems to be 
small in Ref.\cite{CPPACS01}, we estimate the derivative using 
the value of $\chi_S$ at $m=0.1$ and $0.2$.
Using the formula, 
$\partial \langle {\cal O} \rangle / \partial \beta = 
\langle {\cal O} (-\partial S / \partial \beta) \rangle 
-\langle {\cal O} \rangle \langle -\partial S / \partial \beta \rangle, $ 
we obtain 
$\partial (\chi_{\rm S}a^2)/\partial \beta = 1.11(5)$ and $0.82(4)$ at 
$\beta_c$ for $m=0.1$ and $0.2$. Then the second derivative of 
$\epsilon -3p$ is estimated to be
$T^2 \partial^2 ((\epsilon -3p)/T^4)/\partial \mu_{\rm q}^2 = 8.5(1.8)$  
at $m=0.1$, where we use the same value of the beta-function 
as in Sec.~\ref{sec:bcmu}. Finally, 
we obtain $T^2 \partial^2(\epsilon/T^4)/\partial \mu_{\rm q}^2 = 10.6(1.8)$. 
The discrepancy of $\epsilon/T^4$ at the RHIC point from $\mu=0$ 
is about $0.05$.  Once again,
because $\epsilon/T^4\approx6$ at $\beta_c$ for $(m,\mu)=(0.1,0)$
\cite{Kar02}, this is 
a 1\% effect,
suggesting that the $\mu_{\rm q}$-dependence of the equation of state 
is small in the regime of relevance for RHIC.

Next we discuss the relation between the equation of state and 
the phase transition line.
It is of great interest to investigate whether 
the values of the pressure $p(T_c(\mu_{\rm q}),\mu_{\rm q})$ 
and energy density $\epsilon(T_c(\mu_{\rm q}),\mu_{\rm q})$ along the 
transition line are constant or not. 
To this end, consider the line of constant pressure 
in the $(T, \mu_{\rm q})$ plane, i.e., 
\begin{eqnarray}
\Delta p = \frac{\partial p}{\partial T} \Delta T 
+ \frac{\partial p}{\partial (\mu_{\rm q}^2)} \Delta (\mu_{\rm
q}^2) 
= \left[ T^4 \frac{\partial (p/T^4)}{\partial T} + \frac{4p}{T} \right] 
\Delta T 
+ \left[ T^4 \frac{\partial (p/T^4)}{\partial (\mu_{\rm q}^2)} 
\right]
\Delta (\mu_{\rm q}^2) =0, \label{eq:pcons}
\end{eqnarray}
together with a similar relation for $\Delta\epsilon$,
and compare it with the phase transition line.
The slope of the constant pressure
line is then given by
\begin{eqnarray}
\frac{{\rm d} T}{{\rm d} (\mu_{\rm q}^2)} = \left. 
- \frac{\partial(p/T^4)}{\partial (\mu_{\rm q}^2)} 
\right/ \left( \frac{\partial (p/T^4)}{\partial T} + \frac{4p}{T^5} 
\right).
\end{eqnarray}
The derivative $\partial (p/T^4) / \partial T$ can be calculated by 
\begin{eqnarray}
T\frac{\partial (p/T^4)}{\partial T} 
= - \left( \frac{1}{a} \frac{\partial a}{\partial \beta} \right)^{-1} 
\frac{\partial (p/T^4)}{\partial \beta} 
= \left( \frac{1}{a} \frac{\partial a}{\partial \beta} \right)^{-1} 
N_t^4 \left( \left\langle \frac{1}{N_s^3 N_t} \frac{\partial S_g}
{\partial \beta} \right\rangle - \left\langle \frac{1}{N_s^3 N_t} 
\frac{\partial S_g}{\partial \beta} \right\rangle_0 \right),
\end{eqnarray}
where $\langle \cdots \rangle_0$ means the expectation value evaluated
at $T=0$ 
for normalization.
Using the data of Ref.~\cite{Kar00}, $p/T^4=0.27(5)$, 
$\partial (p/T^4) / \partial \beta =4.5(9)$ at $T_c$ for $m=0.1$, 
together with the beta-function in Sec.~\ref{sec:bcmu},
we obtain $T (\partial (p/T^4) / \partial T)\vert_{T=T_c} 
=2.2(6)$ 
for $m=0.1$. 
Noting also that 
$\partial (p/T^4) / \partial (\mu_{\rm q}^2) = 
(1/2) (\partial^2 (p/T^4) / \partial \mu_{\rm q}^2)=0.347(3)/T^2$,
we find that  
the slope of the constant pressure line emerging from the 
critical point on the $T$-axis is 
$T({\rm d} T / {\rm d} (\mu_{\rm q}^2)) = -0.107(22)$.
A similar argument using the data of \cite{Kar02} gives the slope
of the constant energy density line
$T({\rm d} T / {\rm d} (\mu_{\rm q}^2)) = -0.087(23)$.
Because the slope of the transition line in terms of $\mu_{\rm
q}^2$ is 
$T_c ({\rm d} T_c / {\rm d} (\mu_{\rm q}^2)) 
= (1/2) T_c ({\rm d}^2 T_c / {\rm d} \mu_{\rm q}^2) \approx -0.07(3)$,
we deduce that the 
variations of $p$ and $\epsilon$ along the phase transition line
are given by 
\begin{equation}
p(T_c(\mu_{\rm q}),\mu_{\rm q})-p(T_c(0),0)
=\mu_{\rm q}^2T_c^2(0)\times0.12(11)\;\;;\;\;
\epsilon(T_c(\mu_{\rm q}),\mu_{\rm q})-\epsilon(T_c(0),0)=
\mu_{\rm q}^2T_c^2(0)\times1.0(2.2),
\end{equation}
the dominant source of uncertainty in each case being the
location of the phase transition line itself.
Within our errors, therefore, 
both pressure and energy density appear
constant 
along the phase transition line.

\subsection{The phase of the determinant at $\mu \neq 0$}
\label{sec:phase}

Finally we discuss the region of applicability of generic
reweighting approaches.
If the reweighting factor in eqn.~(\ref{eq:rew}) changes 
sign frequently due to the complex phase of the quark determinant, 
then both numerator and denominator of (\ref{eq:rew}) become
vanishingly small in the thermodynamic limit, typically behaving 
$\sim{\rm e}^{-N_{\rm site}}$ with the lattice size 
$N_{\rm site} \equiv N_s^3 N_t$. 
This makes control of statistical errors in
the calculation of the expectation value very difficult. Of course,
$\arg(\det M)$ starts at zero at $\mu=0$ but grows 
as $\mu$ increases. It is important to establish at 
which value of $\mu$ the sign problem becomes severe.

\begin{figure}[t]
\centerline{
\epsfxsize=11.0cm\epsfbox{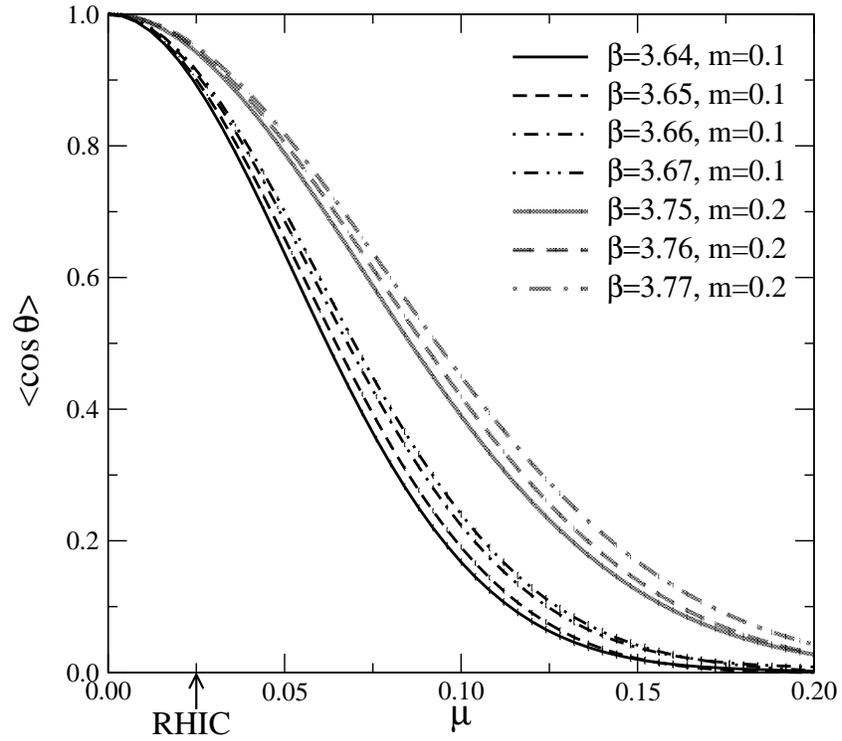}
}
\caption{
The expectation value of the complex phase $\langle \cos \theta \rangle$.
}
\vspace*{-4mm}
\label{fig:phase}
\end{figure}

As discussed in section \ref{sec:method}, the phase can be expressed 
using the odd terms of the Taylor expansion of $\ln \det M$. 
If we write $\det M = |\det M| {\rm e}^{i \theta}$, then
\begin{eqnarray}
\label{eq:phase}
\theta = \alpha N_{\rm f} \ {\rm Im} \left[\mu \frac{\partial \ln \det M}
{\partial \mu} + \frac{\mu^3}{3!} \frac{\partial^3 \ln \det M}{\partial \mu^3} 
+ \cdots \right].
\end{eqnarray}
For small $\mu$, the first term 
$\alpha N_{\rm f} {\rm Im} \, {\rm tr} [M^{-1} (\partial M \partial \mu)] 
\mu$ is dominant. Now, 
because 
$(N_s^3 N_t)^{-1} {\rm tr} [ M^{-1} (\partial M/\partial \mu) ]$ 
is the quark number density, its expectation value must be real and in fact
vanishes
at $\mu=0$. 
Although the average of the phase is zero, 
its fluctuations remain important. 
We investigated the standard deviation of 
$(N_s^3 N_t)^{-1} {\rm Im} \, {\rm tr} [M^{-1} (\partial M/\partial \mu)]$
and present the results in Table \ref{tab:phase}. 
We find values of about $2.2 \times 10^{-3}$ at $\beta_c(m=0.1)$ 
and $1.6 \times 10^{-3}$ at $\beta_c(m=0.2)$. The standard deviation of 
the leading term of (\ref{eq:phase}) therefore has a 
magnitude of about $18 \mu$ 
for $m=0.1$ and $13 \mu$ for $m=0.2$ in the vicinity of the transition.
Consequently the phase problem appears from $\mu \sim 0.09(0.12)$, 
i.e., $\mu_{\rm q}/T_c \sim 0.4(0.5)$ for $m=0.1(0.2)$, 
since the phase problem arises if the phase fluctuation becomes 
of $O(1)$. 
We notice that the value of $\mu$ for which the phase fluctuations become 
significant decreases as either $m$ or $\beta$ decreases. 
Roughly speaking, the numerator and denominator of (\ref{eq:opexp}) 
decrease in proportion to the average of the phase factor 
$\langle {\rm Re}({\rm e}^{i \theta}) \rangle$. 
We show this factor for various $\beta$ and $m$ in Fig.~\ref{fig:phase}, 
where it is clear that the average becomes small around the values of $\mu$
quoted above. The phase fluctuations at the RHIC point $\mu_{\rm q}=0.1T_c$, 
however, are small enough for the analysis of secs.~\ref{sec:bcmu} and
\ref{sec:quarknum} to be applicable. 

We should also note that the fluctuation of the phase depends 
on the lattice size $N_{\rm site}$, 
and on the number of the noise vectors $N_{\rm n}$.
From general arguments, the phase of the reweighting factor is expected to 
decrease as 
$\langle {\rm e}^{i \theta} \rangle \propto {\rm e}^{-N_{\rm site}}$, 
implying that the applicable region of reweighting becomes narrower 
as the lattice size grows. By contrast, the value of 
${\rm Im} \, {\rm tr} [ M^{-1} (\partial M / \partial \mu) ]$ 
calculated on each configuration also contains an error due to the 
finite number of noise vectors (see eqn.(\ref{eq:error}) of the Appendix); 
for $N_{\rm n}=10$ this error is not small compared to 
the standard deviation, 
as seen in Table \ref{tab:phase}. 
The phase fluctuation discussed above includes this error  
due to finite $N_{\rm n}$, and we suspect that the true 
fluctuation becomes smaller as $N_{\rm n}$ increases. To confirm this, 
we reanalyze the standard deviation 
$ \sqrt{\left\langle \{{\rm Im} \, {\rm tr} [M^{-1} 
(\partial M/\partial \mu) ] \}^2 \right\rangle - 
\left\langle {\rm Im} \, {\rm tr} [M^{-1} (\partial M/\partial \mu) ] 
\right\rangle^2}$
by treating the calculation of 
$\left\langle \{{\rm Im} \, {\rm tr} [M^{-1} 
(\partial M/\partial \mu) ] \}^2 \right\rangle$ more carefully. 
Since the noise sets must be independent,
we subtract the contributions from using the same noise vector for each factor. 
Details are given in the Appendix.
The results are quoted in the STD(Imp.) column of Table~\ref{tab:phase}
and are found to be significantly smaller.
Because they might be closer to the $N_{\rm n}=\infty$ limit,
they suggest that the standard deviation for larger $N_{\rm n}$ is 
much smaller, which means that the region of applicability 
becomes wider as $N_{\rm n}$ increases.

\section{Conclusions}
\label{sec:summary}

In this paper we have 
proposed a new method based on a Taylor expansion in chemical potential $\mu$
to investigate the thermodynamic properties of QCD with $\mu\not=0$. 
By computing the chiral susceptibility and the Polyakov loop susceptibility 
for 2 flavors of p-4 improved staggered fermions, we have been able to estimate
the dependence of $\beta_c$, and hence the critical temperature $T_c$, on $\mu$
on moderately large volumes, thus reinforcing the recent advance of lattice QCD
into the interior of the $(\mu_{\rm q},T)$ plane \cite{Hands01}. 
We have also been able to quantify the effect of a non-zero chemical potential
on the equation of state.
Although we have focussed on critical observables in
order to fix physical scales, the method can be applied 
in a small range of $\mu$ at arbitrary $\beta$, although the radius of
convergence is expected to decrease as $T\to0$ since in this limit all
$\mu$-dependence should vanish for $\mu_{\rm q}\leq\mu_o$, making the
behaviour about the origin non-analytic.
The method is also applicable to a 
range of physical observables \cite{Gott88,Gav01,Taro01}.
We find that $T_c$ decreases as $\mu$ increases, but this 
appears to depend only weakly on quark mass, an effect also observed in 
studies of the equation of state $p(T)$ \cite{CPPACS01}.
Our results are in broad agreement with estimates based on exact reweighting 
\cite{Fod01} and
suggest that the discrepancy of $\beta_c$ from its value at
$\mu=0$ is 
small in the interesting region for heavy-ion collisions. 
Moreover we have observed evidence that 
when a negative chemical potential is imposed, 
the generation of dynamical anti-quarks and the consequent screening 
of an external color triplet current is enhanced.

An unresolved issue is the method's limitations. 
We have been able to estimate the complex phase of the fermion determinant 
for a $16^3 \times 4$ lattice and 
found that the sign problem is not serious in the range 
$\mu_{\rm q}/T_c < 0.4$-$0.5$ for $m=0.1$-$0.2$, 
covered by this study.
It is not yet clear to us to
what extent the radius of convergence of the Taylor expansion is linked to the
fluctuations of $\arg(\det M)$. An optimist might hope that the method can 
yield accurate thermodynamic information 
all the way out to the critical
endpoint where the quark/hadron phase transition changes from second to first
order; moreover, since individual terms in the expansion are expectation
values of local operators, the method should be applicable on arbitrarily large
volumes, particularly if larger numbers $N_{\rm n}$ of stochastic noise vectors
than we have used here are employed.
A pessimist might worry that phase
fluctuations should make calculation of higher order terms impracticable
long before the radius of convergence is reached, particularly as the chiral
limit is approached since in this case 
the correlations between $\arg(\det M)$ and $\mbox{Im}({\cal O})$ should 
discriminate between the different
physics associated with isoscalar and isovector chemical potentials.
More work is needed before we can say which is more realistic. 

After this work was submitted we learned of a paper which  
studies the phase transition line by analytical 
continuation of results obtained by simulation with imaginary $\mu$
\cite{DeFPhil}. The results are in reasonable agreement with ours.

\section*{Acknowledgments}
Numerical work was performed using a 128-processor APEmille in Swansea. 
This work was supported by PPARC grant PPA/G/S/1999/00026, 
by the EU contract ERBFMRX-CT97-0122, and by the National
Science Foundation under Grant No. PHY99-07949.


\section*{Appendix: Remark on the noise method}
\label{sec:noise}

The calculation of an operator such as $({\rm tr}A)^2$,
where $A$ is a matrix, using the noise method has to be treated carefully.
Because the random noise vectors should be independent for 
each calculation of ${\rm tr}A$,
\begin{eqnarray}
({\rm tr} A )^2  = \lim_{N_{\rm n} \to \infty} 
\frac{1}{N_{\rm n}} \sum_{a=1}^{N_{\rm n}} \eta^{\dagger}_{a} A \eta_{a} 
\frac{1}{N_{\rm n}} \sum_{b=1}^{N_{\rm n}} \eta^{\dagger}_{b} A \eta_{b} = 
\lim_{N_{\rm n} \to \infty} \frac{1}{N_{\rm n}(N_{\rm n}-1)} \sum_{a \neq b} 
\eta^{\dagger}_{a} A \eta_{a} \eta^{\dagger}_{b} A \eta_{b}.
\end{eqnarray}
This equation can rewritten as
\begin{eqnarray}
\label{eq:no2av}
({\rm tr} A )^2 = \lim_{N_{\rm n} \to \infty} 
\left[ \left( \frac{1}{N_{\rm n}} \sum_{a}^{N_{\rm n}} \eta^{\dagger}_{a} 
A \eta_{a} \right)^2 - \varepsilon^2(A) \right],
\label{eq:error}
\end{eqnarray}
where $\varepsilon (A)$ is the error due to finite $N_{\rm n}$: 
\begin{eqnarray}
\varepsilon^2(A) = \frac{1}{N_{\rm n}-1} \left\{ \frac{1}{N_{\rm n}} 
\sum_{a}^{N_{\rm n}} \left( \eta^{\dagger}_{a} A \eta_{a} \right)^2 
- \left( \frac{1}{N_{\rm n}} \sum_{a}^{N_{\rm n}} 
\eta^{\dagger}_{a} A \eta_{a} \right)^2 \right\}.
\end{eqnarray}
The error 
decreases as $(N_{\rm n}-1)^{-1}$ as $N_{\rm n}$ increases, 
but can be significant for small $N_{\rm n}$.
Moreover, $\varepsilon^2(A)$ is negligible for 
an operator which always has the same sign such as ${\rm tr} M^{-1}$; 
in this case its contribution is about $0.001\%$ for 
$\langle ({\rm tr} M^{-1})^2 \rangle$ with $N_{\rm n}=10$. 
However, for an operator which changes sign frequently such as 
${\rm tr} [M^{-1} (\partial M / \partial \mu)]$, 
the effect of the additional term is important.
We calculate the quark number susceptibility and the value of `STD(Imp.)' 
in Table~\ref{tab:phase} taking this additional term into account.
The difference between `STD' and `STD(Imp.)' in Table~\ref{tab:phase} 
is the contribution from the additional term. 

\begin{figure}[t]
\centerline{
\epsfxsize=11.0cm\epsfbox{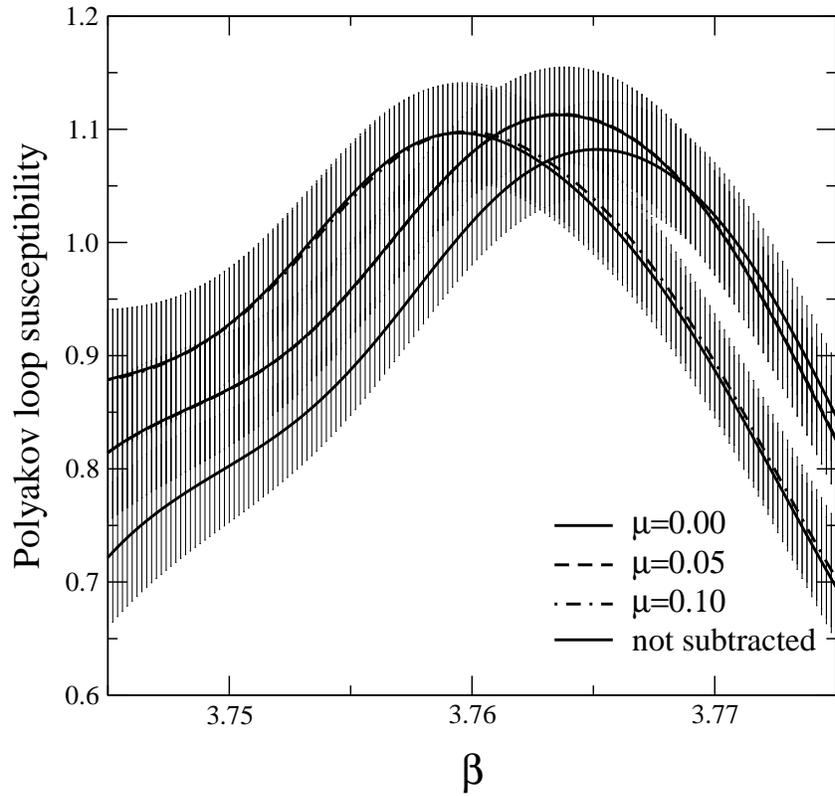}
}
\caption{
Effect from the term of $O(\varepsilon^2)$ on $\chi_L$ 
at $m=0.2$. Solid lines are the same as Fig. \protect\ref{fig:psu02} 
obtained including the $O(\varepsilon^2)$ term, and dashed 
lines are calculated without it.
}
\vspace*{-4mm}
\label{fig:psuno}
\end{figure}

Next, we construct the reweighting method based on Taylor expansion, 
eqn.(\ref{eq:opexp}), explicitly up to second order using the noise 
method. Assuming ${\cal O}$ is a bosonic operator, we can rewrite 
the numerator of eqn.(\ref{eq:opexp}): 
\begin{eqnarray}
&& \left\langle {\cal O} {\rm e}^{\alpha N_{\rm f}
(\ln \det M(m, \mu) - \ln \det M(m_0, 0))}  \right\rangle
= \left\langle {\cal O} \right\rangle 
+ \mu \alpha N_{\rm f} \left\langle {\cal O} {\rm tr} \left(
M^{-1} \frac{\partial M}{\partial \mu} \right) \right\rangle \nonumber \\
&& + \frac{\mu^2}{2} (\alpha N_{\rm f})^2 \left\langle {\cal O} {\rm tr} 
\left(M^{-1} \frac{\partial M}{\partial \mu} \right) {\rm tr} 
\left(M^{-1} \frac{\partial M}{\partial \mu} \right) \right\rangle 
\nonumber \\
&& + \frac{\mu^2}{2} \alpha N_{\rm f} \left[ \left\langle {\cal O} {\rm tr} 
\left(M^{-1} \frac{\partial^2 M}{\partial \mu^2} \right) \right\rangle 
- \left\langle {\cal O} {\rm tr} \left(M^{-1} \frac{\partial M}{\partial \mu} 
M^{-1} \frac{\partial M}{\partial \mu} \right) \right\rangle \right] 
+ \cdots \\
&&= \left\langle {\cal O} \right\rangle 
+ \mu \alpha N_{\rm f} \left\langle {\cal O} 
\overline{\left( \eta^{\dagger}M^{-1} 
\frac{\partial M}{\partial \mu} \eta \right)} \right\rangle \nonumber \\
&& + \frac{\mu^2}{2} (\alpha N_{\rm f})^2 \left[ \left\langle {\cal O} 
\overline{\left(\eta^{\dagger} M^{-1} \frac{\partial M}{\partial \mu} \eta 
\right)}^2 \right\rangle - \left\langle {\cal O} 
\varepsilon^2\left( M^{-1} \frac{\partial M}{\partial \mu}\right) 
\right\rangle \right] \nonumber \\
&& + \frac{\mu^2}{2} \alpha N_{\rm f} \left[ \left\langle {\cal O} 
\overline{\left(\eta^{\dagger} M^{-1} \frac{\partial^2 M}{\partial \mu^2} 
\eta \right)} \right\rangle 
- \left\langle {\cal O} \overline{\left( \eta^{\dagger} M^{-1} 
\frac{\partial M}{\partial \mu} M^{-1} \frac{\partial M}{\partial \mu} 
\eta \right)} \right\rangle \right] + \cdots \\
&&= \left\langle {\cal O} \exp \left\{ 
\mu \alpha N_{\rm f} \overline{\left( \eta^{\dagger}M^{-1} 
\frac{\partial M}{\partial \mu} \eta \right)} \right. \right. \nonumber \\
&& \left. \left. - \frac{\mu^2}{2} (\alpha N_{\rm f})^2 
\varepsilon^2\left( M^{-1} \frac{\partial M}{\partial \mu}\right) 
+ \frac{\mu^2}{2} \alpha N_{\rm f} \left[ 
\overline{\left(\eta^{\dagger} M^{-1} \frac{\partial^2 M}{\partial \mu^2} 
\eta \right)} 
- \overline{\left( \eta^{\dagger} M^{-1} 
\frac{\partial M}{\partial \mu} M^{-1} \frac{\partial M}{\partial \mu} 
\eta \right)} \right] + \cdots \right\} \right\rangle , 
\end{eqnarray}
where $\overline{(\cdots)}$ denotes the average over the noise vectors.
The denominator of eqn.(\ref{eq:opexp}) is given by the same expression with
${\cal O}=1$.
In each case a term proportional to $\varepsilon^2$ appears. 
In Fig.~\ref{fig:psuno}, we estimate the effect of this term by subtracting 
this term from original one. 
The difference for $\chi_L$ by the subtraction is found to be quite small,
e.g., that is less than $1\%$ at $m=0.2$ and $\mu \leq 0.1$.
The result suggests the contribution from the term of $\varepsilon^2$ 
is small for $\chi_L$ although the value of 
$\varepsilon [ M^{-1} (\partial M / \partial \mu)]^2$ 
itself is not small.

For the case of a fermionic operator such as $\bar\psi\psi$
many such additional terms appear in the reweighting formula. 
In this study, we neglect the effect from further additional terms, 
since Fig.~\ref{fig:psuno} suggests that the effect is small for the 
determination of $\beta_c$.


\vspace{1cm}

\begin{table}[tb]
\caption{Simulation point $(m, \beta)$ and number of configurations 
$N_{\rm conf}$ for mass-reweighting and $\mu$-reweighting.
}
\label{tab:para}
\begin{tabular}{cccc}
\hline
  $m$ & $\beta$ & $N_{\rm conf}{\rm (mass)}$ & $N_{\rm conf} (\mu)$ \\
\hline
0.1 & 3.640 & 38000 & 20000 \\
    & 3.645 & 15000 &       \\
    & 3.650 & 58000 & 38000 \\
    & 3.655 & 16800 &       \\
    & 3.660 & 55000 & 40000 \\
    & 3.665 &  7800 &       \\
    & 3.670 & 30000 & 30000 \\
\hline
0.2 & 3.740 &  5000 &       \\
    & 3.750 & 30000 & 20000 \\
    & 3.755 & 15000 &       \\
    & 3.760 & 52000 & 34000 \\
    & 3.770 & 48000 & 32000 \\
    & 3.780 &  5000 &       \\
\hline
\end{tabular}
\end{table}

\begin{table}[tb]
\caption{Quark mass dependence of transition point determined by 
$L$ and $\langle\bar\psi\psi\rangle$.
The fitting function is 
$\beta_c=\beta_c(m_0)+\sum_{n=1}^{N_{\rm fit}}c_i(m-m_0)^n$
The truncation error is contained in $c_2$ from $\bar{\psi}\psi$.}
\label{tab:mass}
\begin{center}
\begin{tabular}{ccccccc}
\hline
  &$m_0$ & $\beta_c(m_0)$ & $c_1$ & $c_2$ & fit range & $N_{\rm fit}$ \\
\hline
$\bar{\psi}\psi$& 0.1 
  & 3.6492(22) & 1.05(14) &      -     & $-0.01 < m-m_0 < 0.01$ & 1 \\
& & 3.6492(22) & 1.03(13) & [$-$9.(14)] & $-0.01 < m-m_0 < 0.01$ & 2 \\
& & 3.6492(22) & 1.07(19) &      -     & $-0.005 < m-m_0 < 0.005$ & 1 \\
& & 3.6492(22) & 1.07(19) & [$-$17.(26)] & $-0.005 < m-m_0 < 0.005$ & 2 \\
& 0.2 
  & 3.7617(36) & 0.896(90)  &    -    & $-0.02 < m-m_0 < 0.02$ & 1 \\
& & 3.7617(36) & 0.894(89)  &  [5.(13)] & $-0.02 < m-m_0 < 0.02$ & 2 \\
& & 3.7617(36) & 0.970(168) &    -    & $-0.01 < m-m_0 < 0.01$ & 1 \\
& & 3.7617(36) & 0.999(180) & [18.(39)] & $-0.01 < m-m_0 < 0.01$ & 2 \\
\hline
Polyakov & 0.2 
  & 3.7639(19) & 0.838(64)  &       -      & $-0.02 < m-m_0 < 0.02$ & 1 \\
& & 3.7639(19) & 0.835(63)  & $-$2.7(4.5)  & $-0.02 < m-m_0 < 0.02$ & 2 \\
& & 3.7639(19) & 0.883(106) &       -      & $-0.01 < m-m_0 < 0.01$ & 1 \\
& & 3.7639(19) & 0.885(106) & $-$4.7(10.0) & $-0.01 < m-m_0 < 0.01$ & 2 \\
\hline
\end{tabular}
\end{center}
\end{table}

\begin{table}[tb]
\caption{$\beta_c$ and its second derivative with respect to $\mu$. 
We fit the data with the function 
$\beta_c (\mu) = \beta_c (0) + \sum_{n=1}^{N_{\rm fit}} c_n \mu^{2n},$ 
where ${\rm d}^2 \beta_c / {\rm d} \mu^2 = 2c_1$.
}
\label{tab:betac}
\begin{center}
\begin{tabular}{ccccccc}
\hline
& $m$ & $\beta_c$ & ${\rm d}^2 \beta_c / {\rm d} \mu^2$ & 
fit range & $N_{\rm fit}$ \\
\hline
$\bar{\psi}\psi$
& 0.1 & 3.6497(16) & $-$1.20(44) & $0 \leq \mu^2 \leq 0.008$ & 1 \\
&     & 3.6497(16) & $-$1.19(54) & $0 \leq \mu^2 \leq 0.005$ & 1 \\
&     & 3.6497(16) & $-$1.21(79) & $0 \leq \mu^2 \leq 0.008$ & 2 \\
& 0.2 & 3.7641(37) & $-$1.02(56) & $0 \leq \mu^2 \leq 0.014$ & 1 \\
&     & 3.7641(37) & $-$1.10(68) & $0 \leq \mu^2 \leq 0.010$ & 1 \\
&     & 3.7641(37) & $-$1.34(103) & $0 \leq \mu^2 \leq 0.014$ & 2 \\
\hline
Polyakov 
& 0.2 & 3.7651(16) & $-$1.01(23) & $0 \leq \mu^2 \leq 0.014$ & 1 \\
&     & 3.7651(16) & $-$1.07(24) & $0 \leq \mu^2 \leq 0.010$ & 1 \\
&     & 3.7651(16) & $-$1.21(31) & $0 \leq \mu^2 \leq 0.014$ & 2 \\
\hline
\end{tabular}
\end{center}
\end{table}

\begin{table}[tb]
\caption{Average of
$\langle {\rm Im} \ {\rm tr} [(\partial M/\partial \mu) M^{-1}] \rangle$, 
average of its error for each configuration $(\langle \varepsilon\rangle)$, 
standard deviation (STD) and improved standard deviation (STD(Imp.)).
}
\label{tab:phase}
\begin{center}
\begin{tabular}{cccccc}
\hline
$m$ & $\beta$ & 
$\langle {\rm Im} \ {\rm tr} [(\partial M/\partial \mu) M^{-1}] \rangle$ &
$\langle \varepsilon \rangle$ & STD & STD(Imp.) \\
\hline
0.1 & 3.64 & $-1.15 \times 10^{-4}$ & 0.00199 & 0.00233 & 0.00110 \\
    & 3.65 & $ 1.02 \times 10^{-5}$ & 0.00194 & 0.00223 & 0.00099 \\
    & 3.66 & $-3.06 \times 10^{-5}$ & 0.00189 & 0.00212 & 0.00085 \\
    & 3.67 & $-1.40 \times 10^{-5}$ & 0.00185 & 0.00206 & 0.00077 \\
\hline
0.2 & 3.75 & $ 1.03 \times 10^{-5}$ & 0.00141 & 0.00168 & 0.00085 \\
    & 3.76 & $ 0.93 \times 10^{-5}$ & 0.00140 & 0.00161 & 0.00072 \\
    & 3.77 & $-4.17 \times 10^{-5}$ & 0.00138 & 0.00155 & 0.00061 \\
\hline
\end{tabular}
\end{center}
\end{table}

\end{document}